\documentclass[onecollarge]{svjour2}       
\smartqed  
\usepackage{psfrag,graphicx,epsfig,color}

\journalname{Journal of Statistical Physics}
\begin{document}

\title{document}
\title{Spherical Ornstein-Uhlenbeck processes}

\author{Michael Wilkinson and Alain Pumir}

\institute{Michael Wilkinson \at
              Department of Mathematics and Statistics, \\
              The Open University,
              Walton Hall, \\
              Milton Keynes, MK7 6AA, \\
              England.\\
              \email{m.wilkinson@open.ac.uk}      \\
             \\
             Alain Pumir \at  
             Laboratoire de Physique,\\
             Ecole Normale Sup\' erieure de Lyon,\\
             F-69007, Lyon,\\
             France\\
             \email{alain.pumir@ens-lyon.fr}}

\date{Received: date / Accepted: date}

\maketitle

\begin{abstract}

The paper considers random motion of a point on the surface
of a sphere, in the case where the angular velocity is determined
by an Ornstein-Uhlenbeck process. The solution is fully characterized
by only one dimensionless number, the persistence angle, which is the
typical angle of rotation of the object during the correlation time of the 
angular velocity.

We first show that the two-dimensional case is exactly
solvable. When the persistence angle is large, a series for the correlation function 
has the surprising property that its sum varies much more slowly than any 
of its individual terms. 

In three dimensions we obtain asymptotic forms for the correlation
function, in the limits where the persistence angle is very small and very
large. The latter case exhibits a complicated transient, followed
by a much slower exponential decay. The decay rate is determined 
by the solution of a radial Schr\" odinger equation in which
the angular momentum quantum number takes an irrational
value, namely $j=\frac{1}{2}(\sqrt{17}-1)$. 
  
Possible applications of the model to objects tumbling in a turbulent
environment are discussed.


\end{abstract}

\section{Introduction}
\label{sec: 1}

There are many contexts in which random motion is confined
to the surface of a sphere. Examples include the motion of a unit vector ${\bf n}(t)$
indicating the orientation of an object tumbling in a turbulent 
fluid flow, or the advection of a tracer in a thin, turbulent planetary
atmosphere. In some applications it is sufficient to model the motion
as diffusion on the surface of a sphere, which can be solved 
by noting that the eigenfunctions of the diffusion operator are 
spherical harmonics. In other applications, however, the angular 
velocity varies smoothly as a function of time, and the diffusive
approximation is not valid. Our objective is to 
obtain insight into random motion of a unit vector, by studying
what is perhaps the simplest model. 

Smooth random motion on a sphere can be characterised by a 
dimensionless parameter which measures the typical  
angular distance through which the point has turned in the timescale
for relaxation of fluctuations of the angular velocity. Due to an analogy with the 
concept of persistence length in polymer physics \cite{degennes+79}, 
we term this parameter
the {\sl persistence angle}: it will be denoted by $\beta$. It is desirable 
to have a model which is a physically well motivated description
of some processes, in which $\beta$ appears as a parameter. 
This paper discusses such a model, which is an extension of the well-known 
Ornstein-Uhlenbeck process \cite{Uhl+30,vKa81} to describe motion on a 
circle or a sphere.

We start by describing the standard Ornstein-Uhlenbeck equation 
on a line, in order to introduce some notation and elementary ideas.
The Ornstein-Uhlenbeck process is a stochastic differential
equation for the time-dependence of a variable $v(t)$:
\begin{equation}
\label{eq: 1.1}
\dot v=-\gamma v+\sqrt{2D}\eta(t)
\end{equation}
where $\eta(t)$ is a white-noise signal, satisfying
\begin{equation}
\label{eq: 1.2}
\langle \eta(t)\rangle=0
\ ,\ \ \ 
\langle \eta(t)\eta(t')\rangle=\delta (t-t')
\ .
\end{equation}
Throughout this paper, $\langle X\rangle$ is the expectation value of $X$. 
The process equilibrates to a statistically stationary state, characterised
by the following correlation function:
\begin{equation}
\label{eq: 1.3}
\langle v(t+\Delta t)v(t)\rangle=\frac{D}{\gamma}\exp(-\gamma|\Delta t|)
\ .
\end{equation}
Equation (\ref{eq: 1.1}) may be considered purely as a model for the fluctuations of velocity $v$ of a particle,
or it may be combined with the equation $\dot x=v$ to give a model for the 
displacement $x$. The motion in space is ballistic when viewed on short timescales,
but on long timescales it is diffusive, with the displacement $\Delta x$ satisfying $\langle \Delta x\rangle=0$ and $\langle \Delta x^2\rangle \sim 2{\cal D}t$, 
with spatial diffusion constant ${\cal D}$:
\begin{equation}
\label{eq: 1.4}
{\cal D}=\frac{1}{2}\int_{-\infty}^\infty {\rm d}t\ \langle v(t)v(0)\rangle=\frac{D}{\gamma^2}
\ .
\end{equation}
To generalize the Ornstein-Uhlenbeck model on a line to a circle, we
simply replace the variable $v$ by the angular velocity, $\omega$, and the
displacement along the line, $x$, by the angle along the circle, $\theta$. In two dimensions, we are 
considering the Ornstein-Uhlenbeck process on a manifold which has the
simplest closed topology, namely a circle. 

If the equation of motion (\ref{eq: 1.1}) is interpreted as a description of an angular 
velocity $\omega$, then $D$ has dimension $[D]={\rm T}^{-3}$. The damping
rate has dimension $[\gamma]={\rm T}^{-1}$. 
Thus, the problem is completely characterized by one single 
dimensionless parameter 
constructed from $D$ and $\gamma$; we take this to be 
\begin{equation}
\label{eq: 1.5}
\beta=\sqrt{\frac{D}{\gamma^3}}
\ .
\end{equation}
The parameter $\beta$ has the following simple physical interpretation. 
The typical angular velocity is, according to (\ref{eq: 1.3}), $\sqrt{D/\gamma}$, and
its fluctuations occur on a timescale $\gamma^{-1}$. The typical angle of rotation over the correlation
timescale of the angular velocity is then $\Delta \theta\sim\sqrt{D/\gamma}/\gamma=\beta$, so that
$\beta$ does correspond to a persistence angle. 

In this paper we show how to compute statistics characterising both circular and spherical Ornstain-Uhlenbeck processes, 
such as the correlation function 
\begin{equation}
\label{eq: 1.6}
C(t)=\langle {\bf n}(t)\cdot {\bf n}(0)\rangle
\ .
\end{equation}
The solution to this problem has very different properties in two and three dimensions.

In two dimensions, exact formulae are obtained for $C(t)$ by expressing this correlation
function in terms of the eigenvalues and eigenfunctions of the Fokker-Planck
equation describing the probability density of the angular velocity and angle variables.
We find that the correlation function $C(t)$ can be expressed as a series: 
\begin{equation}
\label{eq: 1.7}
C(t)=\exp(-\beta^2\gamma t)\sum_{N=0}^\infty C_N(\beta) \exp(-N\gamma t)
\end{equation}
where the coefficients $C_N(\beta)$ are determined explicitly, and satisfy $C_N(0)=\delta_{N,0}$. 
We show that series given by Eq.\ref{eq: 1.7} can be summed exactly, which
reduces the correlation function to a very simple analytic form. There is, however,
a surprising feature of this series expansion which deserves comment.
Although the series (\ref{eq: 1.7}) is convergent for all $t$ and for all $\beta$, in 
the limit as $\beta\to \infty$ the structure of (\ref{eq: 1.7}) appears
to be hard to reconcile with physical expectations. We expect that 
when $\beta\gg 1$, the object rotates with almost constant angular velocity, and the 
decay of correlations results from motions with different angular velocity getting out of
phase. This occurs on a timescale $\sqrt{\gamma/D}=1/(\gamma\beta)$, 
{\sl larger} than the timescale for decay of the leading factor in (\ref{eq: 1.7}),
which is $1/(\gamma \beta^2)$. The implication is that, when $\beta\gg 1$, the summation
over the exponentially {\sl decreasing} terms approaches an {\sl increasing} exponential,
which almost cancels the rapid decay of $\exp(-\beta^2\gamma t)$.
We show how this behaviour is realised, by taking very large coefficients with alternating 
signs. This effect is analogous to a phenomenon known as superoscillation \cite{Ber94}, where
a Fourier sum oscillates faster than appears to be possible due to its bandwidth.
In the case where $\beta\ll 1$, the motion of ${\bf n}(t)$ is well approximated 
by diffusion on the circle, and in that limit (\ref{eq: 1.7}) yields $C(t)=\exp(-Dt/\gamma^2)$, as expected.

The extension of the problem to three-dimensions requires the introduction
of a {\it vector} to describe the angular velocity. In addition, the closed 
manifold that describes the configuration of the system has a significantly
more complicated topology. It turns out that in three-dimensions, the 
Ornstein-Uhlenbeck model of dynamic on a sphere does not appear 
to allow an exact solution, and we show that 
the spectrum has a very different structure from that of the two-dimensional case. When 
$\beta\to 0$, the dynamics is easily understood in terms of diffusion on the surface of a sphere,
for which the eigenfunctions are spherical harmonics. The solution has a much more 
complex behaviour in the limit as $\beta\to \infty$.
As  for the two-dimensional case, there is a transient which occurs on a timescale $1/\gamma\beta$,
but this transient is followed by a slow exponential decay of the correlation function on a timescale
$\sim 1/\gamma$.
We show that in the limit as $\beta\to \infty$ the slowest-decaying mode contributing 
to the correlation function is obtained from a radial Schr\"odinger equation for which
the quantum number $j$ takes an {\sl irrational} value. 
Physically, the two time-scale solution we find for the correlation function
can be explained by noting that when the angular velocity is constant, 
the component of $\bf n$ parallel to $\omega$ is also a constant. The existence 
of this invariance explains the slow 
decorrelation of $\bf n$ at long times. 

Because this problem is far from straightforward we start by considering the two-dimensional case,
where the Ornstein-Uhlenbeck process has a circular coordinate. In section \ref{sec: 2} the 
correlation function (\ref{eq: 1.6}) is determined in closed form for the circular
Ornstein-Uhlenbeck process. The spherical Ornstein-Uhlenbeck process is discussed in section 
\ref{sec: 3}, where we show that the Fokker-Planck operator can be transformed into
a quantum Hamiltonian for a spin coupled to a spherical harmonic oscillator. Section \ref{sec: 3} also
considers the symmtery properties of the Fokker-Planck operator and the 
use of the eigenfunctions of the three-dimensional harmonic oscillator as a convenient basis set.
The asymptotic behaviour of the solution for both large and small $\beta$ is considered 
in section \ref{sec: 4}. Section \ref{sec: 5} contains some discussion of possible areas of
application of the model. Technical details of matrix elements required in section \ref{sec: 4} are 
discussed in appendix A. Appendix B discusses an alternative model for continuous random motion
on a circle, which is required to support the discussion in section \ref{sec: 5}. 

\section{Two-dimensional case}
\label{sec: 2}

\subsection{Formulation and general solution}
\label{sec: 2.1}

We consider motion on a circle, with angular coordinate $\theta$, angular velocity $\omega$,
replacing equation (\ref{eq: 1.1}) by: 
\begin{equation}
\label{eq: 2.1.1}
\dot \theta=\omega
\ ,\ \ \ 
\dot \omega=-\gamma \omega + \sqrt{2D}\eta(t)
\end{equation}
where $\eta(t)$ is a standard white noise signal, with statistics 
satisfying (\ref{eq: 1.2}), and where $\gamma$, $D$
are the damping and diffusion constants. Our aim is to be able to compute 
correlation functions such as $\langle{\bf n}(t)\cdot{\bf n}(0)\rangle$, where
${\bf n}(t)$ is a unit vector with direction $\theta$. This correlation function 
is obtained by computing $\langle \cos\theta (t)\rangle$ subject to the initial
condition $\theta(0)=0$, with $\omega$ having its equilibrium distribution.

Statistics of the model (\ref{eq: 2.1.1}) are obtained by computing the 
joint probability density of $\theta$ and $\omega$, $P(\theta,\omega,t)$,
with an appropriate initial condition. This probability density satisfies a Fokker-Planck
equation \cite{vKa81}
\begin{equation}
\label{eq: 2.1.2}
\frac{\partial P}{\partial t}=-\frac{\partial}{\partial \theta}(\omega P)+\gamma\frac{\partial}{\partial \omega}(\omega P)
+D\frac{\partial^2 P}{\partial \omega^2}\equiv \hat {\cal F}P
\ .
\end{equation}
This equation is solved by determining the spectrum and eigenfunctions of the Fokker-Planck operator
$\hat{\cal F}$, satisfying $\hat{\cal F}\,\Psi_{nm}(\theta,\omega)=\lambda_{nm}\Psi_{nm}(\theta,\omega)$. This operator is separable, and its eigenfunctions are of the  
form $\Psi_{nm}(\theta,\omega)=\exp({\rm i}n\theta)\psi_{nm}(\omega)$. The functions
$\psi_{nm}(\omega)$ satisfy
\begin{equation}
\label{eq: 2.1.3}
\gamma\frac{\partial}{\partial \omega}[\omega \psi_{nm}(\omega)]
+D\frac{\partial^2 }{\partial \omega^2}\psi_{nm}(\omega)-{\rm i}n\omega\psi_{nm}(\omega)
=\lambda_{nm}\psi_{nm}(\omega)
\ .
\end{equation}
To solve this equation, define the operators
\begin{equation}
\label{eq: 2.1.4}
\hat F_0=\gamma \partial_x x+D\partial_x^2
\ ,\ \ \ 
\hat F(\alpha)=\hat F_0-\alpha x
\ .
\end{equation}
The operator $\hat F(\alpha)$ is not Hermitian. It is convenient to introduce the following 
transformation to a Hermitian form:
\begin{equation}
\label{eq: 2.1.5}
\hat H(\alpha)\equiv \exp(\gamma x^2/4D)\hat F(\alpha)\exp(-\gamma x^2/4D)
=D\partial^2_x -\frac{\gamma^2}{4D}x^2 +\frac{\gamma}{2}-\alpha x
\ .
\end{equation}
Note that $\hat H_0\equiv \hat H(0)$ is an inverted harmonic oscillator, with eigenvalues
$-\gamma m$, $m=0,1,\ldots$. It will be convenient to use the Dirac notation for
functions which are acted on by linear operators, replacing the usual angular \lq bra-kets'
with rounded ones to avoid confusion with our notation for expectation values.
Thus the eigenfunctions of $\hat H_0$ are denoted by vectors $|\varphi_m)$:
\begin{equation}
\label{eq: 2.1.6}
\hat H_0\, |\varphi_m)=-\gamma m\,|\varphi_m)
\ .
\end{equation}
The eigenfunctions will be assumed to be normalised according to the quantum
mechanical convention, so that the integral of their modulus squared over all space
is equal to unity, for example the vector $|\varphi_0)$ is a symbolic representation of the 
normalised eigenfunction $(\gamma/2\pi D)^{1/4}\exp(-\gamma x^2/4D)$. These eigenfunctions may be 
generated from $|\varphi_0)$ by repeated application of creation and annihilation
operators, $\hat a^+$ and $\hat a$ respectively. These operators are
\begin{equation}
\label{eq: 2.1.7}
\hat a^+=\sqrt{\frac{D}{\gamma}}\left(\frac{\gamma}{2D}x-\partial_x\right)
\ ,\ \ \ 
\hat a=\sqrt{\frac{D}{\gamma}}\left(\frac{\gamma}{2D}x+\partial_x\right)
\ .
\end{equation}
Note that $\hat a^+$ is the Hermitian conjugate of $\hat a$.
The annihilation and creation operators, $\hat a$, $\hat a^+$ satisfy
\begin{equation}
\label{eq: 2.1.8}
\hat H_0=-\gamma \hat a^+ \hat a
\ ,\ \ \ 
[\hat H_0,\hat a^+]=-\gamma \hat a^+
\ ,\ \ \ 
[\hat H_0,\hat a]=\gamma \hat a
\ ,\ \ \ 
[\hat a,\hat a^+]=1
\ .
\end{equation}
These relations imply that if $\hat H_0\varphi(x)=\lambda\varphi(x)$, then
$\hat a^+\varphi(x)$ and $\hat a\varphi(x)$ are also eigenfunctions,
with eigenvalues $\lambda-\gamma$ and $\lambda+\gamma$ respectively
(with the exception of the ground state, which is destroyed by $\hat a$).
The following relations describe the action of $\hat a$, $\hat a^+$
on normalised eigenfunctions of $\hat H_0$:
\begin{equation}
\label{eq: 2.1.9}
\hat a |\varphi_n)=\sqrt{n}|\varphi_{n-1})
\ ,\ \ \ 
\hat a^+ |\varphi_n)=\sqrt{n+1}|\varphi_{n+1})
\ .
\end{equation}
The eigenfunctions of $\hat H_0$ are orthogonal,
because this operator is Hermitian: 
\begin{equation}
\label{eq: 2.1.10}
(\varphi_n|\varphi_m)\equiv \int_{-\infty}^\infty {\rm d}x\ \varphi_n(x)\varphi_m(x)=\delta_{nm}
\ .
\end{equation}
The eigenfunctions and eigenvalues for other values of $\alpha$ can be
obtained by considering a transformation to a new coordinate $y=x+x_0$,
where the shift is $x_0=2D\alpha/\gamma^2$. In terms of this new coordinate,
we find that $\hat H(\alpha)$ is equivalent to $\hat H_0+D\alpha^2/\gamma^2$.
The eigenvalues $\lambda_m(\alpha)$ and  eigenfunctions $|\varphi_m(\alpha))$ of 
$\hat H(\alpha)$ are therefore
\begin{equation}
\label{eq: 2.1.11}
\lambda_m(\alpha)=-\gamma m +\frac{D\alpha^2}{\gamma^2}
\ ,\ \ \ 
|\varphi_m(\alpha))=\hat T(2D\alpha/\gamma^2)\, |\varphi_m) 
\end{equation}
where $\hat T(X)$ is a translation operator, defined by $\hat T(X)f(x)=f(x-X)$. 
This operator may be represented as an exponential
\begin{equation}
\label{eq: 2.1.12}
\hat T(X)=\exp(-X\partial_x)
\ .
\end{equation}
By expanding the exponential and comparing with the Taylor seriers, we see that 
this expression is consistent with the defining property that $\hat T(X)\,f(x)=f(x-X)$
for any function $f(x)$. Comparing (\ref{eq: 2.1.3}) and (\ref{eq: 2.1.4}), we see that 
we require $\alpha={\rm i}n$, so that we shall require matrix elements of this operator
for complex values of $X$.
 
The eigenfunctions $|\varphi_m)$ of $\hat H_0$ correspond to eigenfunctions $\psi_m(x)$ of $\hat F(\alpha)$, where:
\begin{equation}
\label{eq: 2.1.13}
\psi_m(x)=\exp(-\gamma x^2/4D)\,\hat T(2D\alpha/\gamma^2)\,\varphi_m(x)
\ .
\end{equation}
If a function $f(x)$ is written as a linear combination of these eigenfunctions
\begin{equation}
\label{eq: 2.1.14}
f(x)=\sum_{m=0}^\infty a_m \psi_m(x)
\end{equation}
then, in order to use the orthogonality property (\ref{eq: 2.1.10}), we must multiply $f(x)$ by
$\exp(\gamma x^2/4D)$ and then act on it with $\hat T(-2D\alpha/\gamma^2)$, to obtain
\begin{equation}
\label{eq: 2.1.15}
a_m=\int_{-\infty}^\infty {\rm d}x\ \varphi_m(x)\,\hat  T(-2D\alpha/\gamma^2)\, \exp(\gamma x^2/4D)\, f(x)
\ .
\end{equation}
In order to evaluate these coefficients, it will also be useful to have an expression for matrix elements 
of the translation operator (Franck-Condon factors) in the basis of
the harmonic oscillator eigenfunctions: 
\begin{equation}
\label{eq: 2.1.16}
I_{nm}(X)\equiv (\varphi_n|\hat T(X)|\varphi_m)\equiv \int_{-\infty}^\infty {\rm d}x\ \varphi_n(x)\varphi_m(x-X) 
\ .
\end{equation}
By a simple adaptation of an argument presented in \cite{Wil87}, for $n\ge m$ these 
can be shown to be given by
\begin{equation}
\label{eq: 2.1.17}
I_{nm}(X)=\sqrt{\frac{m!}{n!}}\left(\frac{X}{2}\sqrt{\frac{\gamma}{D}}\right)^{n-m}
\exp\left(-\frac{X^2\gamma}{8D}\right)L_{m}^{(n-m)}\left(\frac{X^2\gamma}{4D}\right)
\end{equation}
where $L^{(\alpha)}_{N}(x)$ is the associated Laguerre polynomial \cite{Abr+72}:
\begin{equation}
\label{eq: 2.1.18}
L^{(\alpha)}_N(x)=\sum_{k=0}^N \frac{(-1)^k}{k!} 
\left(\begin{array}{c}
n+\alpha \cr n-k
\end{array}\right) 
x^k
\ . 
\end{equation}
The case $n\le m$ is obtained by using the fact that $I_{mn}(X)=I_{nm}(-X)$. 

Using these results and noting that comparison of (\ref{eq: 2.1.3}) and (\ref{eq: 2.1.4}) 
implies $\alpha={\rm i}n$, we see
that the eigenfunctions of the Fokker-Planck operator $\hat {\cal F}$
are 
\begin{equation}
\label{eq: 2.1.19}
\psi_{nm}(\theta,\omega)=\exp({\rm i}n\theta)\exp(-\gamma \omega^2/4D)\hat T(2{\rm i}nD/\gamma^2)\varphi_m(\omega)
\ .
\end{equation}
These eigenfunctions are complex-valued for $n\ne 0$, but note that the degenerate 
eigenfunctions $\psi_{n,m}$ and $\psi_{-n,m}$ are complex conjugates, so that we can
form real-valued solutions of (\ref{eq: 2.1.2}). A general solution of (\ref{eq: 2.1.2}) can be written in the form
\begin{equation}
\label{eq: 2.1.20}
P(\theta,\omega,t)=\sum_{n=-\infty}^\infty \sum_{m=0}^\infty a_{nm}\, \exp[-\gamma(m+n^2D/\gamma^3)t]
\, \psi_{nm}(\theta,\omega)
\end{equation}
where the coefficients $a_{nm}$ are determined by the initial conditions:
using (\ref{eq: 2.1.15}) gives
\begin{equation}
\label{eq: 2.1.21}
a_{nm}=\frac{1}{2\pi}\int_0^{2\pi}{\rm d}\theta\ \exp(-{\rm i}n\theta) \int_{-\infty}^\infty{\rm d}\omega\ 
\varphi_m(\omega)\,\hat T(-2{\rm i}nD/\gamma^2)\,\exp(\gamma \omega^2/4D)\,P(\theta,\omega,0)
\ .
\end{equation}

\subsection{Evaluation of correlation functions}
\label{sec: 2.2}

\begin{figure}
\centering
\begin{tabular}{cc}
\epsfig{file=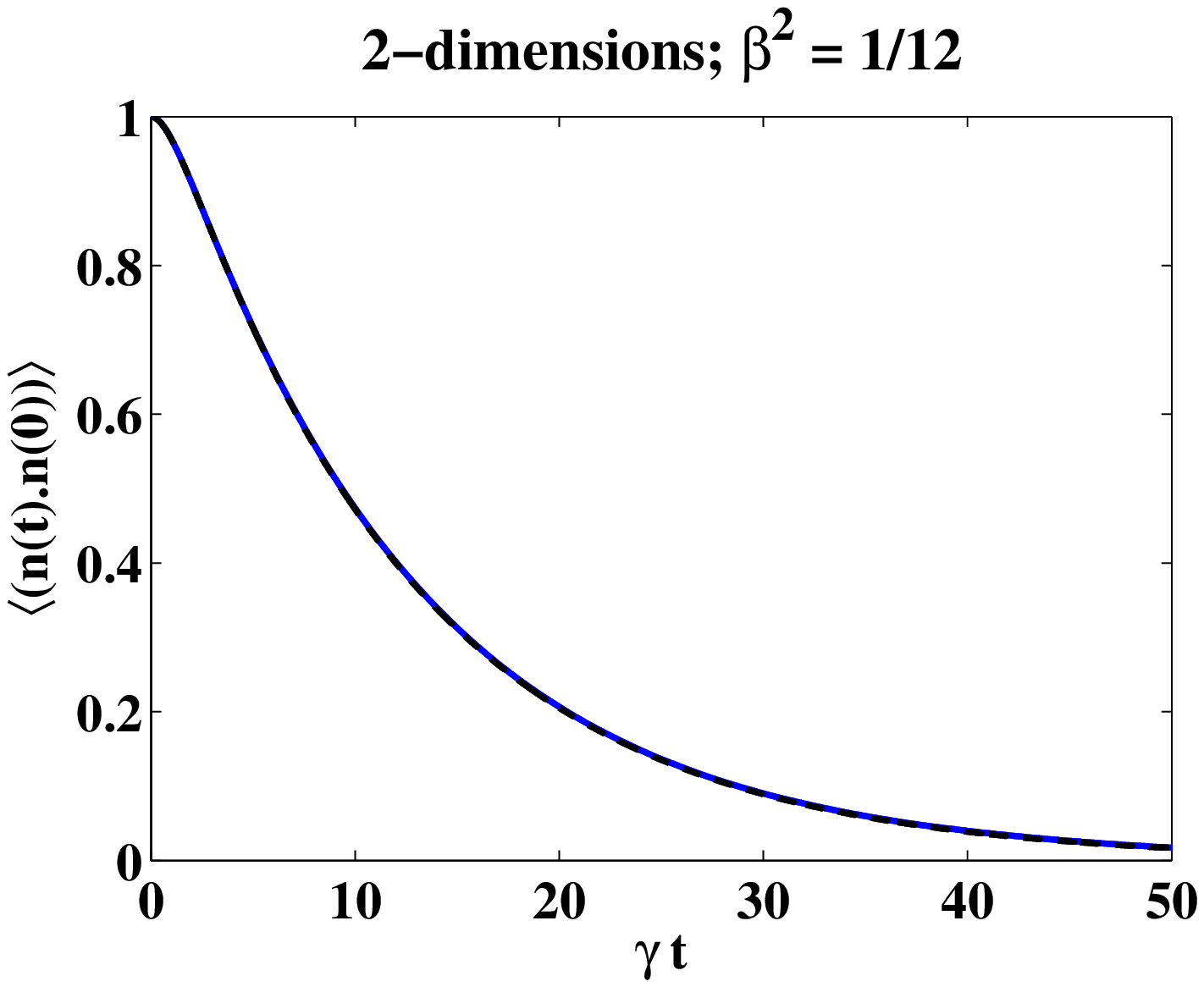,width=0.4\linewidth,clip=} & 
\epsfig{file=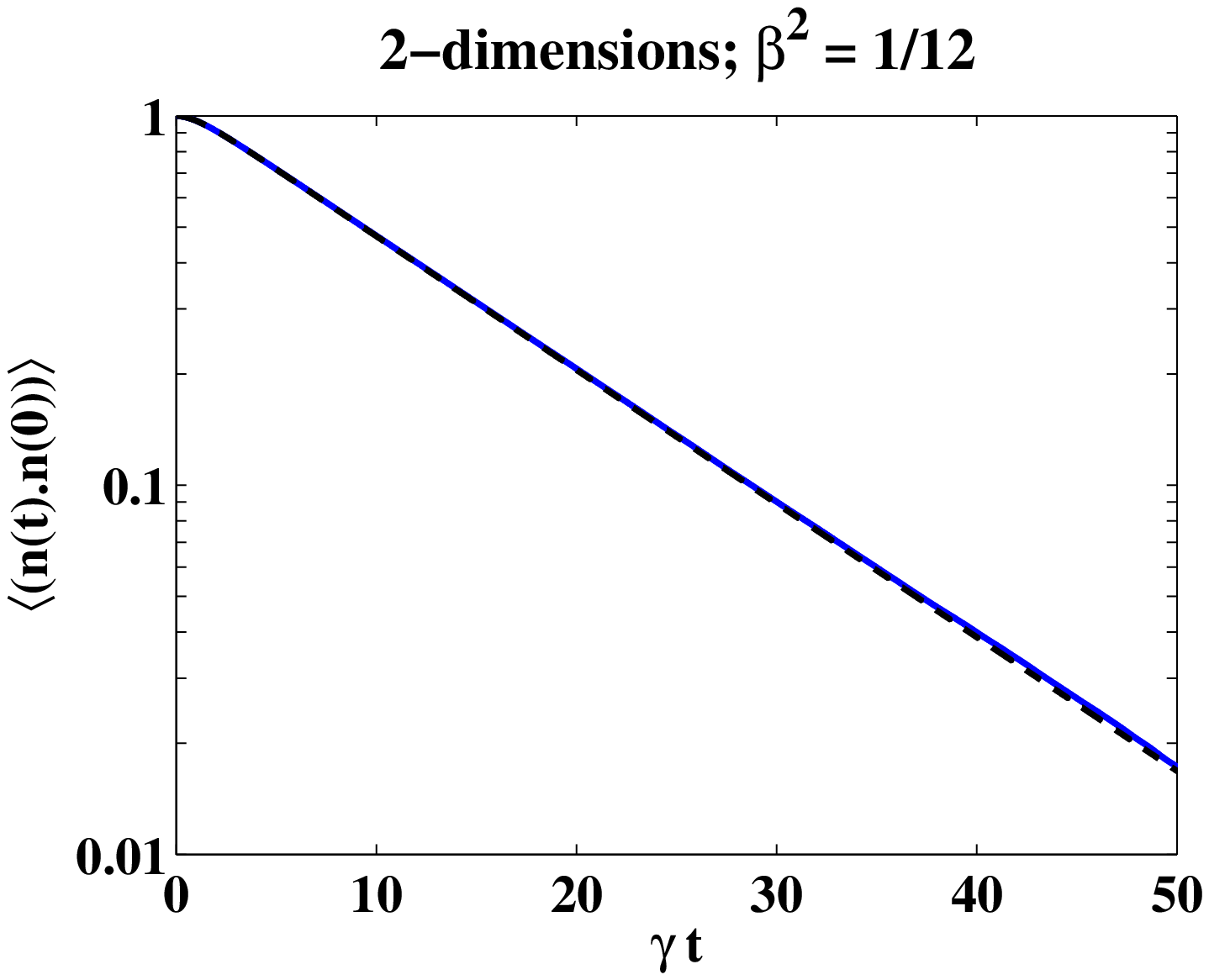,width=0.4\linewidth,clip=} \\
\epsfig{file=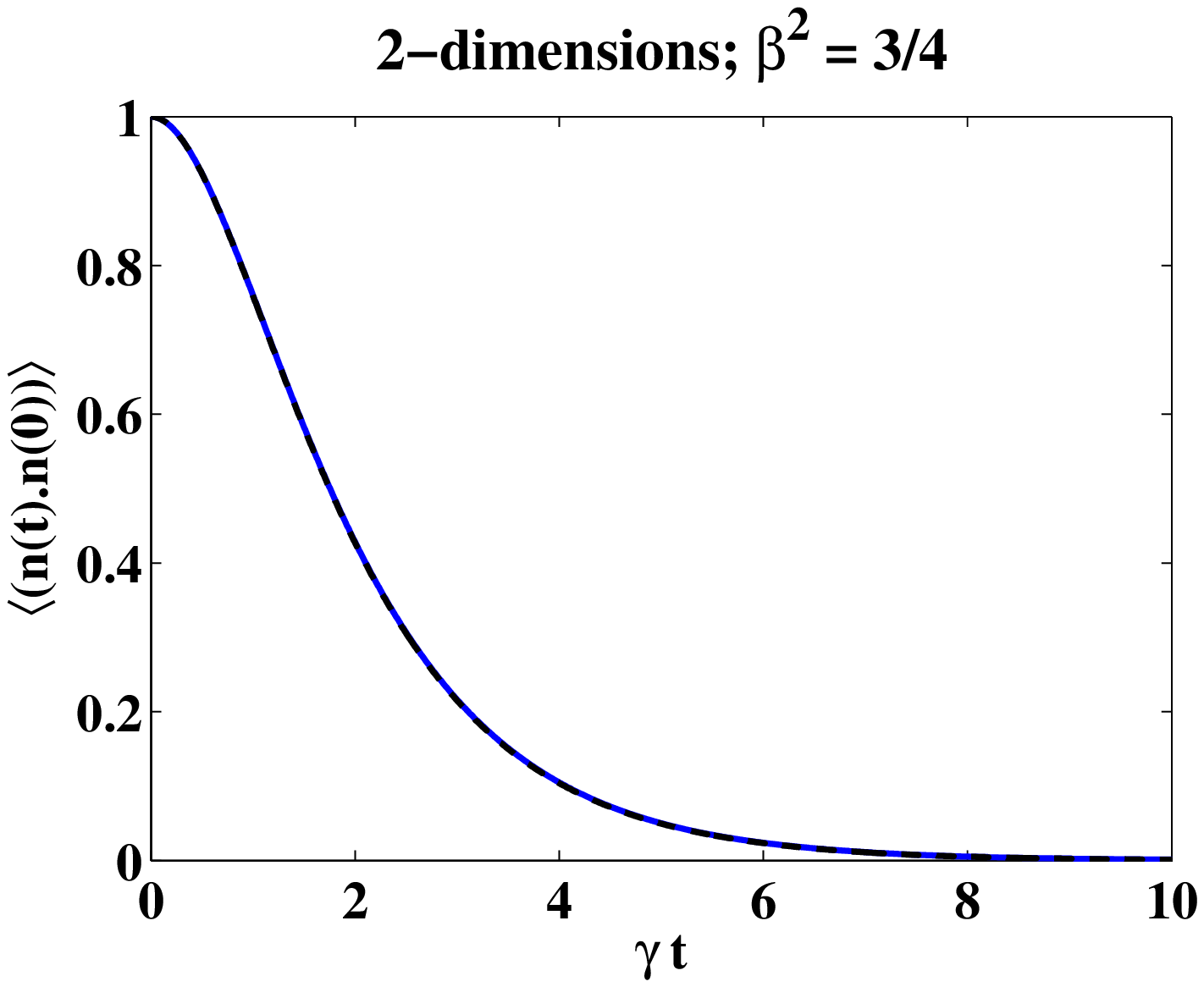,width=0.4\linewidth,clip=} &
\epsfig{file=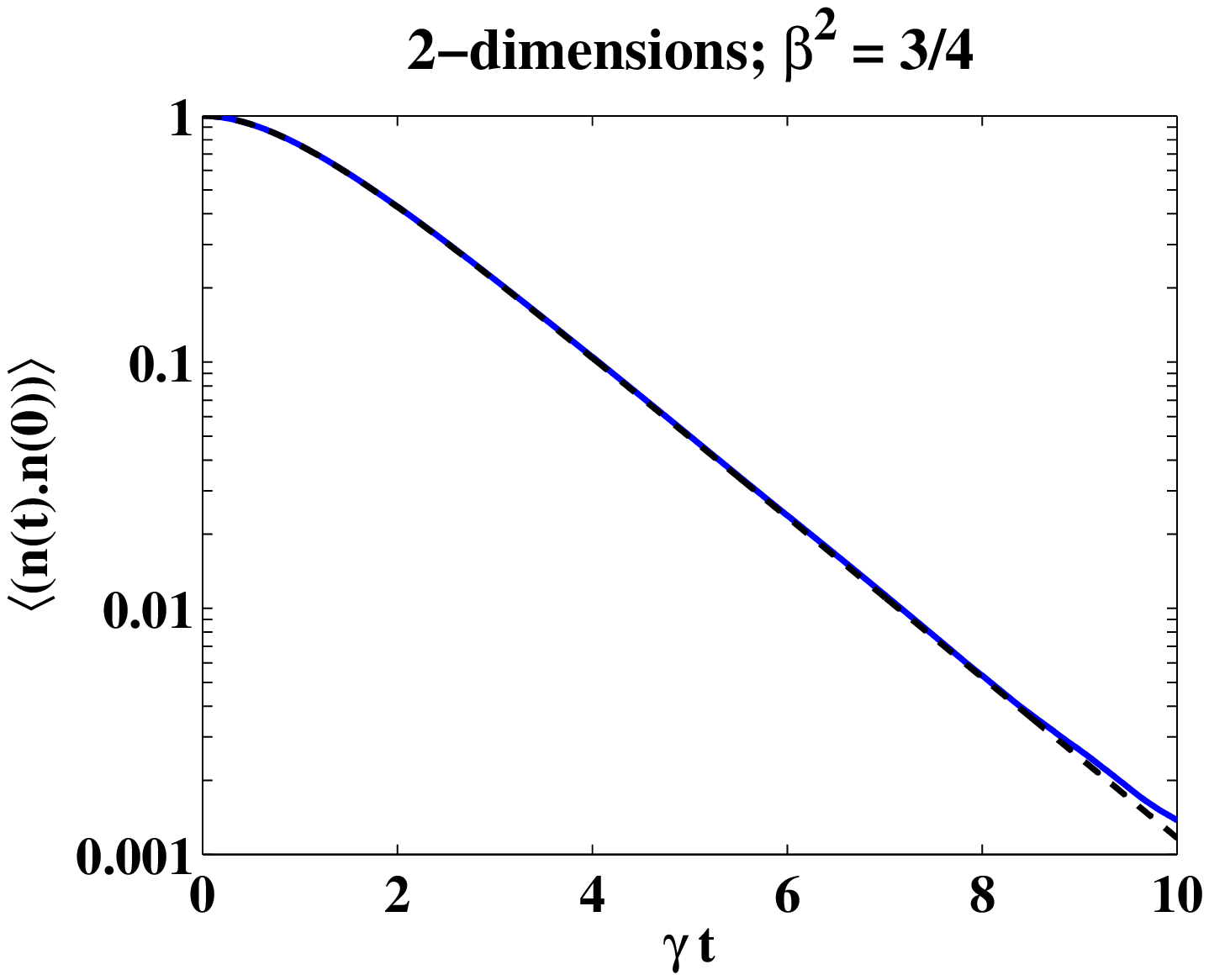,width=0.4\linewidth,clip=}\\
\epsfig{file=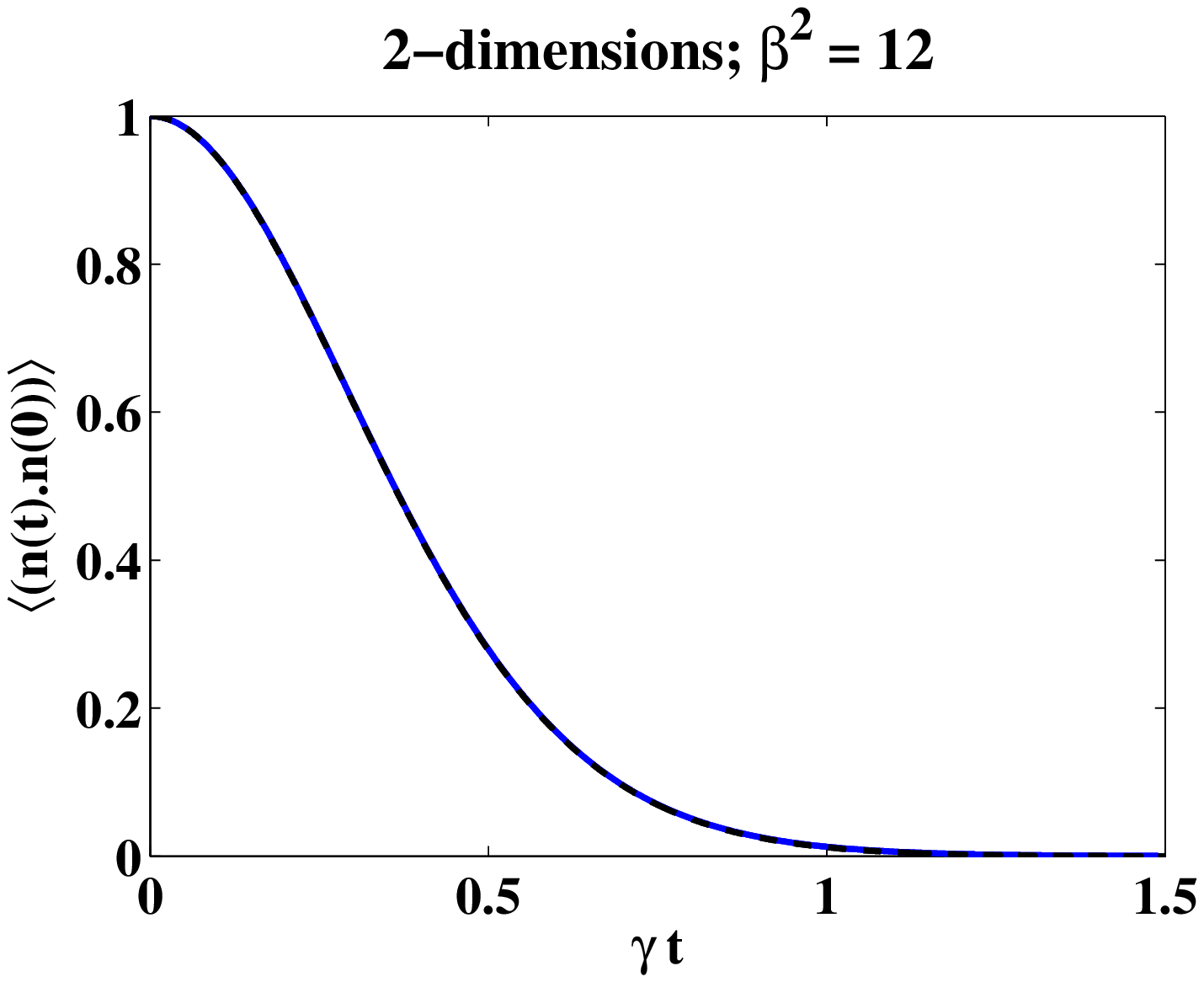,width=0.4\linewidth,clip=} &
\epsfig{file=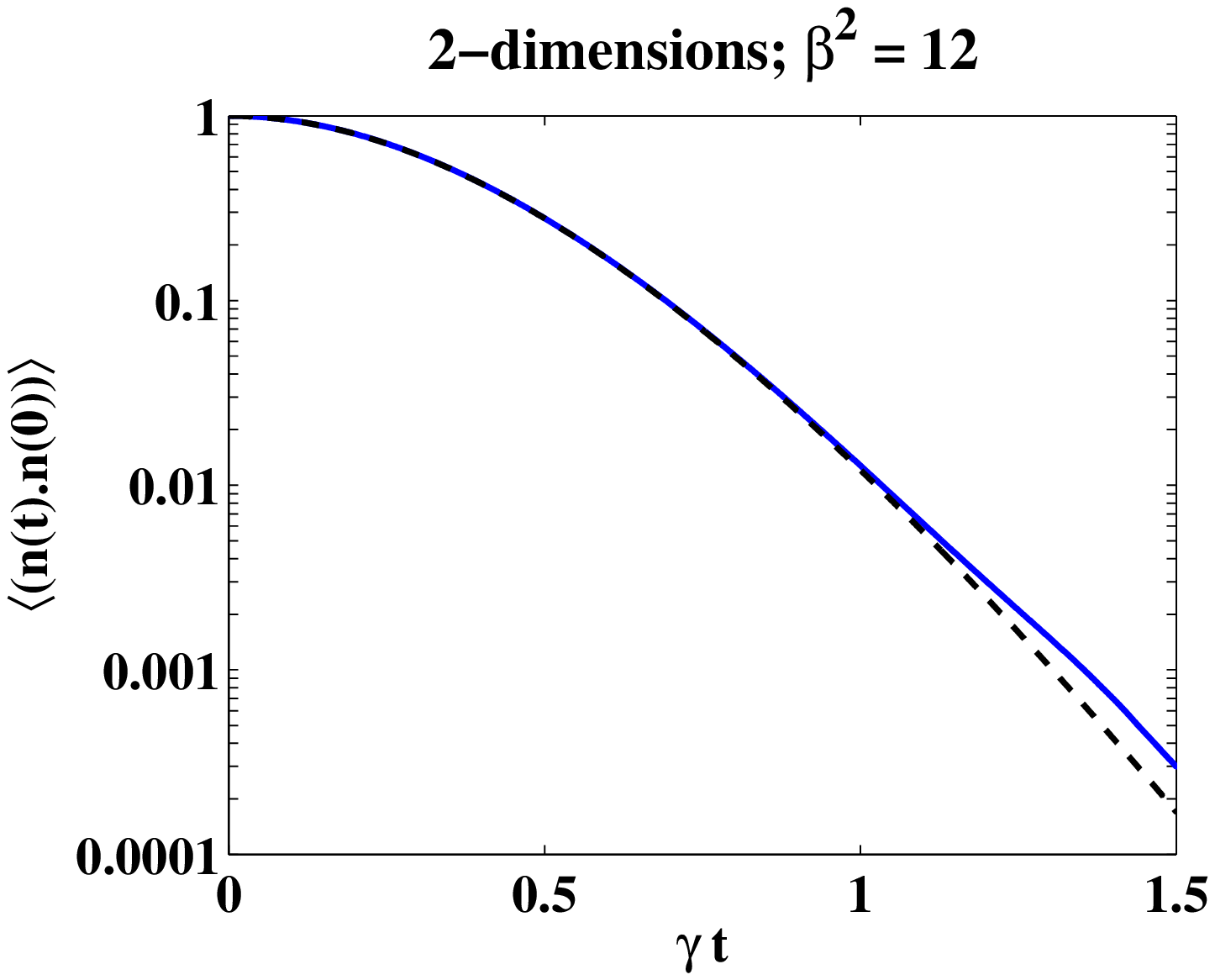,width=0.4\linewidth,clip=}
\end{tabular}
\caption{\label{fig: 1} Correlation function for the circular Ornstein-Uhlenbeck
process, for three values of $\beta^2=D/\gamma^3$: 
$\beta^2 = 1/12$ (upper row), $\beta^2 = 3/4$ (middle row) and $\beta^2 = 12$ (lower row). 
The simulations (full curves) show 
excellent agreement with the theoretical results, equation (\ref{eq: 2.2.7}), 
shown as a dashed line.}
\end{figure}

In order to evaluate the correlation function $C(t)$ defined by (\ref{eq: 1.6}), we 
consider the distribution of $\theta$ with the initial 
condition that the initial orientation ($\theta=0$, say) is known, but the angular momentum
distribution initially in equilibrium:
\begin{equation}
\label{eq: 2.2.1}
P(\theta,\omega,0)=\sqrt{\frac{\gamma}{2\pi D}}\exp(-\gamma \omega^2/2D)\,\delta(\theta)=\left(\frac{\gamma}{2\pi D}\right)^{1/4}\delta(\theta)\,
\exp(-\gamma\omega^2/4D)\,\varphi_0(\omega)
\ .
\end{equation}
Using (\ref{eq: 2.1.21}), the coefficients in (\ref{eq: 2.1.20}) are
\begin{equation}
\label{eq: 2.2.2}
a_{nm}=\frac{1}{2\pi}\left(\frac{\gamma}{2\pi D}\right)^{1/4}\int_{-\infty}^\infty {\rm d}\omega\ \varphi_m(\omega)\,\hat T(-2{\rm i}nD/\gamma^2)\,\varphi_0(\omega)
=\frac{1}{2\pi}\left(\frac{\gamma}{2\pi D}\right)^{1/4}I_{m0}(-2{\rm i}nD/\gamma^2)
\ .
\end{equation}
The correlation function is
\begin{eqnarray}
\label{eq: 2.2.3}
C(t)&=&\langle {\bf n}(t)\cdot {\bf n}(0)\rangle=\langle \cos \theta \rangle
\nonumber \\
&=&\int_0^{2\pi}{\rm d}\theta\ \cos \theta \int_{-\infty}^\infty {\rm d}\omega\ P(\theta,\omega,t)
\nonumber \\
 &=&2\pi \exp(-Dt/\gamma^2)\sum_{m=0}^\infty {\rm Re}\left[a_{1m}\ \exp(-\gamma m t) \int_{-\infty}^\infty {\rm d}\omega\ 
\exp(-\gamma \omega^2/4D)\,\hat T(2{\rm i}D/\gamma^2)\,\varphi_m(\omega)\right]
\nonumber \\
&=&\exp(-Dt/\gamma^2)\sum_{m=0}^\infty I_{m0}(-2{\rm i}D/\gamma^2)\, I_{0m}(2{\rm i}D/\gamma^2) \,\exp(-\gamma m t)
\ .
\end{eqnarray}
Note that (\ref{eq: 2.1.17}) implies that the only associated Laguerre polynomials which are 
required are $L_0^{(\alpha)}(x)=1$, so that the Franck-Condon factors in (\ref{eq: 2.2.3}) are
\begin{equation}
\label{eq: 2.2.4}
I_{m0}(2{\rm i}D/\gamma^2)=\sqrt{\frac{1}{m!}}(-{\rm i}\beta)^m\,\exp(\beta^2/2)=I_{0m}(-2{\rm i}D/\gamma^2)
\end{equation}
where $\beta=\sqrt{D/\gamma^3}$ is the dimensionless parameter defined in (\ref{eq: 1.5}), 
so that the correlation function in (\ref{eq: 2.2.3}) is
\begin{equation}
\label{eq: 2.2.5}
C(t)=\exp(-Dt/\gamma^2)\exp(\beta^2)\sum_{m=0}^\infty \frac{(-\beta^2)^m}{m!}\exp(-m\gamma t)
\ .
\end{equation}
Using the identity
\begin{equation}
\label{eq: 2.2.6}
\sum_{m=0}^\infty \frac{(-x)^m}{m!}\exp(-am)=\exp[-x\exp(-a)]
\end{equation}
the correlation function can be obtained is closed form:
\begin{equation}
\label{eq: 2.2.7}
C(t)=c(\gamma t,\sqrt{D/\gamma^3})
\ ,\ \ \ \ 
c(\tau,\beta)=\exp[\beta^2(1-\tau-\exp(-\tau))]
\ .
\end{equation}
Figure \ref{fig: 1} shows a comparison between the correlation 
function obtained by numerical averaging for the circular Ornstein-Uhlenbeck
process, for three different values of $\beta$. In each case the results are compared
with the theoretical expression, equation (\ref{eq: 2.2.7}), and the agreement 
is excellent. 

\subsection{Discussion}
\label{sec: 2.3}

In the limits $\beta \to 0$ and $\beta \to \infty$ the correlation function approaches 
limiting forms which are exponential and Gaussian, respectively: from (\ref{eq: 2.2.7})
we find
\begin{eqnarray}
\label{eq: 2.3.1}
C(t)&\sim &\exp(-Dt/\gamma^2)\ ,\ \ \ \beta\ll 1
\\
\label{eq: 2.3.2}
C(t)&\sim &\exp(-Dt^2/2\gamma)\ ,\ \ \ \beta \gg 1
\ .
\end{eqnarray}
It is instructive to consider how these limiting cases arise. 
In the case where $\sqrt{D/\gamma^3}\ll 1$, the partial
probability density $P(\theta,t)$ satisfies a diffusion equation
\begin{equation}
\label{eq: 2.3.3}
\frac{\partial P}{\partial t}={\cal D}\frac{\partial^2 P}{\partial \theta^2}
\end{equation}
which is to be solved with the initial condition $P(\theta,0)=\delta(\theta)$.
The solution of this equation on the circle is
\begin{equation}
\label{eq: 2.3.4}
P(\theta,t)=\sum_{m=-\infty}^\infty a_m \exp({\rm i}m\theta)\exp(-m^2{\cal D}t)
\ .
\end{equation}
The initial condition gives $a_m=1/2\pi$, so that 
\begin{eqnarray}
\label{eq: 2.3.5}
\langle \cos\theta (t)\rangle&=&\frac{1}{2\pi}\sum_{m=-\infty}^\infty \exp(-m^2{\cal D}t
)\int_0^{2\pi}{\rm d}\theta\ \cos(\theta) \exp({\rm i}m\theta)
\nonumber \\
&=&\exp(-{\cal D}t)=\exp(-Dt/\gamma^2)
\end{eqnarray}
in agreement with (\ref{eq: 2.3.1}).
In the opposite limit, $\sqrt{D/\gamma^3}\gg 1$, the particle rotates around the circle 
at a rate which is equal to its initial angular velocity. The equilibrium distribution of 
angular momentum has density 
\begin{equation}
\label{eq: 2.3.6}
P(\omega)=\sqrt{\frac{\gamma}{2\pi D}}\exp(-\omega^2\gamma /2D)
\ .
\end{equation}
After time $t$ the angle is $\theta=\omega t$, so that the probability distribution
of the angle is
\begin{equation}
\label{eq: 2.3.7}
P(\theta,t)=\sqrt{\frac{\gamma}{2\pi Dt^2}}\exp(-\theta^2 \gamma /2Dt^2)
\ .
\end{equation}
The correlation function is then
\begin{eqnarray}
\label{eq: 2.3.8}
\langle \cos\theta(t)\rangle&=&\sqrt{\frac{\gamma}{2\pi Dt^2}}\int_{-\infty}^\infty {\rm d}\theta\ 
\cos(\theta)\exp(-\theta^2\gamma/2Dt^2)
\nonumber \\
&=&\exp(-Dt^2/2\gamma)
\end{eqnarray}
in agreement with (\ref{eq: 2.3.2}).

The introduction mentioned that there is a surprising aspect to the behaviour 
of the series (\ref{eq: 2.2.5}) in the limit as $\beta \to \infty$. Despite the fact 
that every term in this series has an exponential decay with a timescale shorter 
than $1/\gamma\beta^2$, the exact evaluation of the sum of this series, 
equation (\ref{eq: 2.2.7}), decays on a much longer timescale,
$1/(\gamma\beta)$.

\section{Three dimensional case}
\label{sec: 3}

\subsection{Formulation and Fokker-Planck equation}
\label{sec: 3.1}

We consider a unit vector ${\bf n}(t)$ evolving according to the equation:
\begin{equation}
\label{eq: 3.1.1}
\frac{{\rm d}{\bf n}}{{\rm d}t}= \mbox{\boldmath$\omega$} \wedge {\bf n}
\end{equation}
where $\mbox{\boldmath$\omega$}$ is the angular velocity vector.
The components $\omega_i$ of the angular velocity are determined
by independent Ornstein-Uhlenbeck equations:
\begin{equation}
\label{eq: 3.1.2}
\frac{{\rm d}\omega_i}{{\rm d}t}=-\gamma \omega_i+\sqrt{2D}\eta_i(t)
\end{equation}
where the $\eta_i(t)$ are independent white noise signals, with statistics
specified by equation (\ref{eq: 1.2}). This process
is described by a Fokker-Planck equation for the joint probability density
of ${\bf n}$ and $\mbox{\boldmath$\omega$}$. The general form for 
the Fokker-Planck
equation in a space with coordinates $x_i$ is \cite{vKa81}:
\begin{equation}
\label{eq: 3.1.3}
\frac{\partial P}{\partial t}=-\sum_{i=1}^6 \frac{\partial}{\partial x_i}(v_i P)
+\sum_{i=1}^6 \sum_{j=1}^6 \frac{\partial^2}{\partial x_i\partial x_j}(D_{ij}P)
\ .
\end{equation}
The velocities $v_i$ and diffusion coefficients $D_{ij}$ are defined 
in terms of the expectation values of increments $\delta x_i$ in time $\delta t$
by writing $\langle \delta x_i\rangle=v_i\delta t$ and $\langle \delta x_i\delta x_j\rangle=2D_{ij}\delta t$.
In our case $(x_1,x_2,\ldots,x_6)=\mbox{\boldmath$x$}=({\bf n},\mbox{\boldmath$\omega$})=(n_1,n_2,n_3,\omega_1,\omega_2,\omega_3)$.
The velocities are
\begin{equation}
\label{eq: 3.1.4}
v_i=\epsilon_{ijk}\omega_j n_k\ ,\ \ \ i=1,2,3\ ,\ \ \ v_i=-\gamma \omega_i
\ ,\ \ \ i=4,5,6 
\end{equation}
and the diffusion coefficients are
\begin{equation}
\label{eq: 3.1.5}
D_{ij}=D\delta_{ij}\ ,\ \ \ i=4,5,6\ ,\ \ \ D_{ij}=0\ ,\ \ \ i=1,2,3
\ .
\end{equation}
The Fokker-Planck equation is therefore
\begin{equation}
\label{eq: 3.1.6}
\frac{\partial P}{\partial t}=\hat {\cal F}P=\epsilon_{ijk}\frac{\partial}{\partial n_i}(n_j\omega_kP)+\gamma \frac{\partial}{\partial \omega_i}(\omega_i P)+
D\frac{\partial^2 P}{\partial \omega_i\partial \omega_i}
\end{equation}
where in equation (\ref{eq: 3.1.6}), as well as in the equations below,
repeated indices are summed over the values $1,2,3$.
The Fokker-Planck operator can also be expressed in the form
\begin{equation}
\label{eq: 3.1.7}
\hat {\cal F}=\gamma \partial_i \omega_i+D\partial_i\partial_i-\omega_i \hat J_i
\end{equation}
where $\hat J_i$ are components of an angular momentum operator, defined by
\begin{equation}
\label{eq: 3.1.8}
\hat J_i=\epsilon_{ijk}n_j\frac{\partial}{\partial n_k}
\end{equation}
and where $\partial_i=\partial/\partial \omega_i$. Note that this definition differs from that
which is commonly used in quantum mechaincs texts (such as \cite{Lan+58}) by a factor of ${\rm i}=\sqrt{-1}$. 
The Fokker-Planck operator (\ref{eq: 3.1.7}) can also be expressed in the form:
\begin{equation}
\label{eq: 3.1.9}
\hat {\cal F}=\hat{\cal F}_0-\mbox{\boldmath$\omega$}\cdot \hat{\mbox{\boldmath$J$}}
\ ,\ \ \ 
 \hat {\cal F}_0=\gamma \mbox{\boldmath$\nabla$}\cdot \mbox{\boldmath$\omega$}
+D\mbox{\boldmath$\nabla$}\cdot\mbox{\boldmath$\nabla$}
\ .
\end{equation}
where $\mbox{\boldmath$\nabla$}$ is the gradient in the angular momentum space.

The variables may be made dimensionless by using:
\begin{eqnarray}
\label{eq: 3.1.10}
\bar t & \equiv & \gamma t  \\
 \bar \omega_i & \equiv & \sqrt{\frac{\gamma}{2D}}\omega_i 
\ . 
\label{eq: 3.1.10b}
\end{eqnarray}

In dimensionless form the Fokker-Planck equation reads:
\begin{equation}
\label{eq: 3.1.11}
\partial_t P = \partial_i (\omega_i P)+\sqrt{2}\beta\epsilon_{ijk} \omega_k \partial_{n_i}( n_j P) + 
\frac{1}{2}\partial^2_i P
\end{equation}
where the variables in equation (\ref{eq: 3.1.11}) are dimensionless, as defined in 
equations (\ref{eq: 3.1.10},\ref{eq: 3.1.10b}); for simplicity, the overbars have been 
omitted.

\subsection{Symmetry analysis}
\label{sec: 3.2}

Consider the symmetry properties of the Ornstein-Uhlenbeck operator
describing the evolution of the joint probability density function 
$P({\bf n},\mbox{\boldmath$\omega$}, t)$.
The physical properties of the system are invariant under the rotation 
group, in the sense that
an arbitrary rotation of both 
${\bf n}$ and $\mbox{\boldmath$\omega$}$ leaves the problem unchanged. 
We define angular 
momentum operators which generate these rotations: 
\begin{eqnarray}
\label{eq: 3.2.1}
\hat L_i & = &  \epsilon_{ijk} \omega_j \partial_{\omega_k}
\nonumber \\
\hat J_i & = &  \epsilon_{ijk} n_j \partial_{n_k}
\end{eqnarray}
(the operators $\hat J_i$ were already considered in (\ref{eq: 3.1.8})).
These operators are related to the operators used in quantum mechanics by a 
factor ${\rm i}$, that is 
$\hat{\mbox{\boldmath$L$}} ={\rm  i} \hat{\mbox{\boldmath$L$}}_{\rm QM}$
where $\hat{\mbox{\boldmath$L$}}_{\rm QM}$ is the angular momentum
operator defined in standard texts such as \cite{Lan+58}. 
Thus, adapting standard results \cite{Lan+58}, the commutation relations of the operators are:
\begin{equation}
\label{eq: 3.2.3}
[\hat L_i,  \hat L_j ] = - \epsilon_{ijk} \hat L_k
\end{equation}
and similar relations for $\hat J_i$. It is also useful to note that:
\begin{equation}
\label{eq: 3.2.4}
[\hat L_i , \omega_j] = - \epsilon_{ijk} \omega_k
\ .
\end{equation}
The term $\mbox{\boldmath$\omega$} \cdot \hat{\mbox{\boldmath$J$}}$ is invariant when rotating {\it simultaneously}
${\bf n}$ and $\mbox{\boldmath$\omega$}$, and using (\ref{eq: 3.2.3}) and (\ref{eq: 3.2.4}) it is easy to check that:
\begin{equation}
\label{eq: 3.2.5}
[\hat{\mbox{\boldmath$J$}} + \hat{\mbox{\boldmath$L$}}, \mbox{\boldmath$\omega$} \cdot \hat{\mbox{\boldmath$J$}}] = 0
\ .
\end{equation}
In the same way, it is easy to see that
\begin{equation}
\label{eq: 3.2.6}
[\hat {\mbox{\boldmath$J$}}^2 , \mbox{\boldmath$\omega$}\cdot \hat{\mbox{\boldmath$J$}} ] = 0
\ .
\end{equation}
It is clear that these are also symmetries of the other elements of the Fokker-Planck operator in (\ref{eq: 3.1.11}).  
Because the system is invariant under rotation simultaneously of $\mbox{\boldmath$\omega$}$ and ${\bf n}$, the Fokker-Planck evolution operator commutes with the {\it total} angular momentum, as well as with the magnitude 
of the angular momentum of ${\bf n}(t)$. The set of conserved quantities is therefore
\begin{equation}
\label{eq: 3.2.7}
\hat{\mbox{\boldmath$K$}}=\hat{\mbox{\boldmath$L$}}
+\hat{\mbox{\boldmath$J$}}
\ ,\ \ \ \
\hat{\mbox{\boldmath$J$}}^2
\end{equation}
where $\hat{\mbox{\boldmath$L$}}$ acts on the $\mbox{\boldmath$\omega$}$ variables and $\hat{\mbox{\boldmath$J$}}$ acts on the ${\bf n}$ variables. The corresponding quantum numbers are denoted $j,m_j$ for
$\hat{\mbox{\boldmath$J$}}$, $l,m_l$ for $\hat{\mbox{\boldmath$L$}}$, and $k,m_k$ for
$\hat{\mbox{\boldmath$K$}}$. 

Thus, the Fokker-Planck operator has a block-diagonal structure, where the blocks are 
labelled by a given set of  values of $j$, $k$ and $m_k$, where $-j(j+1)$ 
is the eigenvalue of $\hat{\mbox{\boldmath$J$}}^2$, $m_k$ the 
eigenvalue of $\hat{\mbox{\boldmath$K$}}_z$ and $-k(k+1)$ the eigenvalue
of $\hat{\mbox{\boldmath$K$}}^2$.
These symmetries indicate that spherical polar coordinates and spherical harmonic functions will prove useful.
We use $(\theta,\phi)$ as the polar angles for ${\bf n}$ and $(\omega,\theta',\phi')$ as the spherical
polar coordinates of $\mbox{\boldmath$\omega$}$. The spherical harmonic functions $Y_{lm}(\theta,\phi)$
will be denoted by Dirac state vectors, writing $|Y_{lm})$ when the arguments are $(\theta,\phi)$ 
and $|Y'_{lm})$ when the arguments are $(\theta',\phi')$.

Further constraints due to symmetry considerations can follow from the initial conditions.
For example, the initial condition for evaluation of correlation functions  
has the variable $\mbox{\boldmath$\omega$}$ in the equilibrium state, which is spherically 
symmetric with $l = 0$ (and which is easily seen to be a Gaussian function of $\mbox{\boldmath$\omega$}$). 
The value of ${\bf n}$ is set equal to one particular vector, say
${\bf e}_z$. The initial condition is therefore
\begin{equation}
\label{eq: 3.2.8}
P({\bf n},\mbox{\boldmath$\omega$}, 0) =\sqrt{4\pi}\left(\frac{\gamma}{2\pi D}\right)^{3/2} 
\exp(-\gamma \omega^2/2D) \ Y_{00}(\theta',\phi') \ \delta({\bf n} - {\bf e}_z )
\ .
\end{equation}
The $\delta$-function distribution can be resolved into a sum the spherical harmonics, 
with $j=0,1,2,\dots$ but with $m_j=0$ in each case:
\begin{equation}
\label{eq: 3.2.9}
\delta({\bf n} - {\bf e}_z) = \sum_{j=0}^{\infty} \frac{2j+1}{4 \pi} 
P_j({\bf n}\cdot {\bf e}_z )
\end{equation}
where the functions $P_j(x)$ are Legendre polynomials.
This means that we can confine attention
to the $m_k=0$ subspace when evaluating an equilibrium correlation 
function of ${\bf n}(t)$.

Furthermore, in the case of the simplest correlation function, $C(t)=\langle \cos \theta\rangle$, 
the quanitity being averaged spans only the $j=1$ subspace, so we can confine
our attention to the $j=1$ subspace.
Thus in order to evaluate $C(t)$ we must consider the $j=1$, $m_k=0$ subspace.
Furthermore, the initial condition has $l=0$, so the total angular
momentum is $k=1$. The values of $\mbox{\boldmath$K$}^2$ and $\mbox{\boldmath$J$}^2$ are 
constants of the motion, indicating that we must consider only solutions with $k=1$ and $j=1$. 
Setting $k=1$ and $j=1$, the triangle relation for $\mbox{\boldmath$L$}=\mbox{\boldmath$K$}-\mbox{\boldmath$J$}$ 
is $|k-j|\le l \le k+j$, that is, $l=0,1,2$.

We can construct functions of the angular variables with definite values
of $\mbox{\boldmath$J$}^2$, $\mbox{\boldmath$L$}^2$, $\mbox{\boldmath$K$}^2$ and
$K_z$, labelled by quantum numbers  $j$, $l$, $k$, $m_k$. Let these functions be
denoted by $\Upsilon_{j,l,k,m_k}(\theta,\phi,\theta',\phi')$, and we assume that these 
functions are normalised so that
they form an orthonormal set, with the usual integration measure for a cartesian product 
of two spherical surfaces.  The symmetry considerations discussed above imply that the
solution in the $j=1$ subspace may be written in the form
\begin{equation}
\label{eq: 3.2.10}
P_1({\bf n},\mbox{\boldmath$\omega$},t)=\sum_{l=0}^2 \psi_l(\omega,t)\ \Upsilon_{1,l,1,0}(\theta,\phi,\theta',\phi')
\ .
\end{equation}
This result shows that, if we are concerned with evaluating the correlation function 
$C(t)=\langle \cos\theta\rangle$, then symmetry considerations reduce the Fokker-Planck 
equation to a system of three coupled ordinary differential equations. 
More generally, the calculation of correlation functions such as 
$\langle P_j(\cos\theta ) \rangle$ implies values of $l$ in the range 
$ 0 \le l \le 2j$, so the solution requires $2j + 1$ functions of $\omega$.

\subsection{Harmonic oscillator basis}
\label{sec: 3.3}

It will also be useful to consider the exact solution of the Ornstein-Uhlenbeck 
process describing the evolution of the angular momentum $\mbox{\boldmath$\omega$}$, 
independent of evolution of ${\bf n}$. This will be related to the three-dimensional quantum 
spherical harmonic oscillator.

The operator $\hat {\cal F}_0$ in (\ref{eq: 3.1.9}) has a structure which is closely related to the harmonic oscillator
of quantum mechanics. It is convenient to transform the Fokker-Planck operator $\hat {\cal F}_0$ 
into a three-dimensional isotropic harmonic oscillator. We consider the operator
\begin{equation}
\label{eq: 3.3.1}
\hat {\cal H}=\exp(\chi/2)\hat {\cal F}\exp(-\chi/2)
\ ,\ \ \ 
\chi=\frac{\gamma(\omega_1^2+\omega_2^2+\omega_3^2)}{2D}
\ .
\end{equation}
Using the dimensionless variables defined by (\ref{eq: 3.1.10}) and (\ref{eq: 3.1.10b}),
we can express $\hat {\cal H}$ in the form
\begin{equation}
\label{eq: 3.3.2}
\hat {\cal H}=-\gamma \left[\hat {\mbox{\boldmath$a$}}^+\cdot
\hat{\mbox{\boldmath$a$}}
+\sqrt{2}\beta \hat{\mbox{\boldmath$J$}}\cdot \mbox{\boldmath$\omega$}\right]\equiv \hat {\cal H}_0-\sqrt{2}\beta\gamma
\hat {\mbox{\boldmath$J$}}\cdot \mbox{\boldmath$\omega$}
\end{equation}
where the components of $\hat {\mbox{\boldmath$a$}}^+$ and $\hat{\mbox{\boldmath$a$}}$ are,
respectively, creation and annilhilation operators for the $i$ degree of 
freedom, using the dimensionless variables defined by equations (\ref{eq: 3.1.10},
\ref{eq: 3.1.10b}) :
\begin{equation}
\label{eq: 3.3.3}
\hat a^+_i=\frac{1}{\sqrt{2}}\left(\omega_i-\partial_i\right)
\ ,\ \ \ \ 
\hat a_i=\frac{1}{\sqrt{2}}\left(\omega_i+\partial_i\right)
\ .
\end{equation}
The eigenvalues of $\hat{\cal H}_0$ are $-\gamma(k_1+k_2+k_3)$, where $k_i=0,1,2,\ldots$. Note that all of
the eigenvalues except the ground state are degenerate. 

As well as being separable in Cartesian coordinates, $\hat {\cal H}_0$ is
also separable in spherical polar coordinates:
\begin{equation}
\label{eq: 3.3.harm_osc}
\hat{\cal H}_0 = \gamma \left[\frac{1}{2} \nabla_\omega^2 - \frac{1}{2} \omega^2 + \frac{3}{2}\right]
\end{equation}
 where the Laplacian operator 
may be expressed as
\begin{equation}
\label{eq: 3.3.4}
\nabla_\omega^2  \equiv \partial_\omega^2 + \frac{2}{\omega} \partial_\omega  
+ \frac{1}{\omega^2}\hat {\mbox{\boldmath$L$}}^2
\end{equation}
and $\hat{\mbox{\boldmath$L$}}$ is, up to a factor ${\rm i}$, 
the usual angular momentum operator acting on 
${\bf e}_\omega$; its eigenvalues are $-l (l + 1)$, $l=0,1,2,\ldots $.
The degenerate multiplets can, therefore,  also
be resolved as states which are eigenfunctions of $\hat{\mbox{\boldmath$L$}}^2$ and $\hat L_z$, labelled
by quantum numbers $n,l,m$ (where $-l(l+1)$ and $m$ are eigenvalues 
of $\hat {\mbox{\boldmath$L$}}^2$ and $\hat L_z$ respectively). The ground state 
eigenfunction of $\hat{\cal H}_0=-\gamma \hat{\mbox{\boldmath$a$}}^+\cdot \hat{\mbox{\boldmath$a$}}$ is 
\begin{equation}
\label{eq: 3.3.5}
\varphi_{000}(\mbox{\boldmath$\omega$})=\left(\frac{1}{2\pi }\right)^{3/4}\exp\left(-\frac{\omega^2}{2}\right)
\end{equation}
and the other eigenfunctions of $\hat{\cal H}_0$ are of the form

\begin{equation}
\label{eq: 3.3.6}
\varphi_{nlm}(\mbox{\boldmath$\omega$})= \sqrt{4\pi}Y_{lm}(\theta',\phi') \,\omega^l P_{nl}\left(\omega^2\right)
\varphi_{000}(\mbox{\boldmath$\omega$})
\end{equation}

where $P_{nl}(x)$ is a polynomial of degree $n$, proportional to the 
generalized Laguerre polynomial $L_n^{(l + 1/2)}$ \cite{Lan+58}. 
The eigenfunctions are normalised in the usual way: 
\begin{eqnarray}
\label{eq: 3.3.7}
\delta_{nn'}&=&
\int_0^\infty {\rm d}x\ x^{2+2l} \exp(-x^2)P_{nl}(x^2)P_{n'l}(x^2)
\nonumber \\
&=&\frac{1}{2} \int_0^\infty {\rm d}y\ y^{l+1/2}\exp(-y)P_{nl}(y)P_{n'l}(y)
\ .
\end{eqnarray}
Using the orthonormality relation for the generalized Laguerre polynomials:
\begin{equation}
\label{eq: 3.3.8}
\int_0^\infty {\rm d}x \ x^\alpha \exp(-x) L^{(\alpha)}_n(x) L^{(\alpha)}_m(x)=\delta_{nm}
\frac{\Gamma (n+\alpha+1)}{n!}
\ .
\end{equation}
we deduce the relation between $P_{nl}$ and the generalized Laguerre
polynomials: 
%
\begin{equation}
\label{eq: 3.3.9}
P_{nl}(x)= {\cal N}_{nl} ~ L^{(l + \frac{1}{2})}_n(x)\ , \ \ \ \ \ 
{\cal N}_{nl} = \sqrt{\frac{2 n!}{(n+l+\frac{1}{2})\Gamma(n+l+\frac{1}{2})}}
\ .
\end{equation}

\subsection{Equations of motion for $j=1$ modes}
\label{sec: 4.3}

Here we consider how to write an equation of motion for the projection onto the $j=1$ modes, 
which contribute to the correlation function $C(t)$. The harmonic oscillator basis 
which was introduced in section \ref{sec: 3.3} will prove useful here.

In section \ref{sec: 3.2} we showed how symmetry considerations constrain the angular
dependences of the solutions. The angle-dependent parts of the solution are constructed
from functions with known values of $\mbox{\boldmath$J$}^2$, $\mbox{\boldmath$L$}^2$, $\mbox{\boldmath$K$}^2$ 
and $K_z$, with quantum numbers $j,l,k,m_k$. These functions may be expressed 
in terms of tensor products of spherical harmonics, writing 
\begin{equation}
\label{eq: 4.3.1}
|\Upsilon_{j,l,k,m_k})=\sum_{l_1,m_1}\sum_{l_2,m_2} (l_1,m_1; l_2,m_2|j,l,k,m_k) |Y_{l_1,m_1})\otimes |Y'_{l_2,m_2})
\end{equation}
where $|Y_{l_1,m_1})$ represents the spherical harmonic $Y_{l_1,m_1}(\theta,\phi)$ which is a 
function of polar angles representing the direction of ${\bf n}$, and $|Y'_{l_2,m_2})$ represents
$Y_{l_2,m_2}(\theta'.\phi')$, which is a function of the polar angles for $\mbox{\boldmath$\omega$}$.
The coefficients $(l_1,m_1;l_2,m_2|j,l,k,m_k)$ are termed Clebsch-Gordon coefficients 
(see, for example, \cite{Edm57,Lan+58}), and this representation
is useful because the spherical harmonics have well-known and convenient properties.

In particular, determining the correlation function $C(t)$ requires a solution involving just three 
functions $\psi_l(\omega,t)$, multiplying the angular functions $\Upsilon_{1,l,1,0}$, with $l=0,1,2$ 
(see equation (\ref{eq: 3.2.10})).
From tabulations of Clebsch-Gordon coefficients we find:
\begin{eqnarray}
\label{eq: 4.3.2}
| \Upsilon_{1, 0, 1,0})& = & | Y_{10}) \otimes | Y'_{00} )
\nonumber \\
| \Upsilon_{1, 1, 1,0}) & = & 
\sqrt{\frac{1}{2}} |Y_{1,+1}) \otimes |Y'_{ 1,-1})  -\sqrt{\frac{1}{2}} | Y_{1,-1}) \otimes | Y'_{1,+1})
\nonumber \\
| \Upsilon_{1,2, 1,0}) & = & 
\sqrt{\frac{3}{10}} | Y_{2,1}) \otimes | Y'_{1,-1}) 
- \sqrt{\frac{2}{5}} | Y_{2,0})\otimes | Y'_{1,0}) 
+ \sqrt{\frac{3}{10}} | Y_{2,-1}) \otimes | Y'_{1,+1}) 
\ .
\end{eqnarray}
In order to express the equation of motion in the $j=1$ subspace, it is necessary to rewrite the operator 
$\mbox{\boldmath$\omega$} \cdot \mbox{\boldmath$J$}$ so that its action upon the spherical 
harmonics is explicit. To this end, consider the angular momentum ladder operators: 
\begin{equation}
\label{eq: 4.3.3}
\hat J_+ = ( \hat J_x + {\rm i} \hat J_y)
\ ,\ \ \ \  
\hat J_- = ( \hat J_x - {\rm i} \hat J_y) 
\end{equation}
It is straightforward to see that:
\begin{equation}
\label{eq: 4.3.4}
[ \hat J_z,  \hat J_{\pm} ] =  \pm {\rm i} \hat J_{\pm}
\ .
\end{equation}
Thus, applying the operator $\hat J_+$ to the eigenstate of $\hat J_z$ with 
a quantum number $m$, namely the spherical harmonic $|Y_{ l, m})$, one finds  
$\hat J_z \hat J_\pm |Y_{ l, m})  ={\rm i} (m \pm 1 ) \hat J_\pm | Y_{l, m})$.
The operator $\hat J_+$ thus increases the azimuthal quantum number by one, whereas 
$\hat J_-$ decreases the azimuthal quantum number by one. 
For completeness, the prefactor, up to a phase, can be obtained by 
expressing $\hat{\mbox{\boldmath$J$}}^2$ as:
\begin{equation}
\label{eq: 4.3.5}
\hat {\mbox{\boldmath$J$}}^2 = \hat J_+ \hat J_- + \hat J_z^2 - {\rm i}\hat J_z =\hat  J_- \hat J_+ + {\hat J_z}^2 +{\rm i} \hat J_z
\end{equation}
which immediately leads to:
\begin{eqnarray}
\label{eq: 4.3.6}
\hat J_+ | Y_{l, m-1}) & = &{\rm i} \sqrt{ ( l+m) (l-m+ 1) }|Y_{ l, m}) 
\nonumber \\
\hat J_- | Y_{l, m}) & = &{\rm i} \sqrt{ ( l+m) (l-m+ 1) } | Y_{l, m-1})
\ .
\end{eqnarray}
With these results in place, we can express $\mbox{\boldmath$\omega$}\cdot \hat{\mbox{\boldmath$J$}}$ in terms
of the operators $\hat J_{\pm}$. Elementary algebra leads to:
\begin{equation}
\label{eq: 4.3.7}
\mbox{\boldmath$\omega$} \cdot \hat {\mbox{\boldmath$J$}} = 
\frac{1}{2} (\omega_x - i \omega_y ) \hat J_+
+ \frac{1}{2} (\omega_x + i \omega_y ) \hat J_-
+ \omega_z \hat J_z
\ .
\end{equation}
If one notices further that:
\begin{eqnarray}
\label{eq: 4.3.8}
(\omega_x + {\rm i} \omega_y)  & = & - \omega \sqrt{\frac{8 \pi}{3} } Y_{1,1} ( \theta', \phi' ) 
\nonumber \\
(\omega_x - {\rm i} \omega_y)  & = & + \omega \sqrt{\frac{8 \pi}{3} } Y_{1,-1} ( \theta', \phi' ) 
\nonumber \\
\omega_z   & = & w \sqrt{\frac{4 \pi}{3} } Y_{1,0} ( \theta', \phi' )
\end{eqnarray}
then, equation (\ref{eq: 4.3.7}) leads to:
\begin{equation}
\label{eq: 4.3.9}
\mbox{\boldmath$\omega$} \cdot \mbox{\boldmath$J$} = \omega \sqrt{\frac{4 \pi}{3}} \Bigl( \frac{1}{\sqrt{2}} 
Y_{1,-1} (\theta', \phi') \hat J_+
+ Y_{1,0} (\theta', \phi') \hat J_z
-  \frac{1}{\sqrt{2}} Y_{1,1} (\theta', \phi') \hat J_- \Bigr)
\ .
\end{equation}
The formulation of equation (\ref{eq: 4.3.9}) is useful to understand how 
the operator $\mbox{\boldmath$\omega$} \cdot \hat{\mbox{\boldmath$J$}}$
couples modes to each other. In particular, with the help of equations 
(\ref{eq: 4.3.9}) and (\ref{eq: 4.3.2}) we can determine the matrix elements
\begin{equation}
\label{eq: 4.3.10}
(\Upsilon_{1,i,1,0}|\mbox{\boldmath$\omega$}\cdot \hat{\mbox{\boldmath$J$}}|\Upsilon_{1,j,1,0})
\equiv \omega {\cal A}_{ij}
\end{equation}
where $i,j\in \{ 0,1,2 \}$.
Only four of these matrix elements are non-zero. After a lengthy but mechanical calculation we 
find the following values for the non-zero matrix elements:
\begin{eqnarray}
\label{eq: 4.3.11}
{\cal A}_{01}=\sqrt{\frac{2}{3}}&\ \ \ \ \ &{\cal A}_{10}=-\sqrt{\frac{2}{3}}
\nonumber\\
{\cal A}_{12}=\sqrt{\frac{1}{3}}&\ \ \ \ \ &{\cal A}_{21}=-\sqrt{\frac{1}{3}}
\ .
\end{eqnarray}
It is convenient to use the connection with the spherical harmonic 
oscillator, and to replace the functions $\psi_l(\omega,t)$ in (\ref{eq: 3.2.10}) by
\begin{equation}
\label{eq: 4.3.12}
\zeta_l(\omega,t)=\exp(-\omega^2/4)\,\psi_l(\omega,t)
\end{equation}
(here we use the dimensionless variables (\ref{eq: 3.1.10}), (\ref{eq: 3.1.10b})).
Substituting (\ref{eq: 3.2.10}), (\ref{eq: 4.3.12}) into the 
Fokker-Planck equation, multiplying by $|\Upsilon_{1,j,1,0})$, 
and integrating over the product of two spheres, we obtain three partial differential equations
for the three components coupling to the $j=1$ mode. These equations are:
\begin{eqnarray}
\label{eq: 4.3.13}
\frac{\partial}{\partial t} \zeta_0(\omega,t)&=& \hat L_0 \zeta_0(\omega,t) +\sqrt{2}\beta \omega  {\cal A}_{01} \zeta_1(\omega,t) 
\nonumber \\
\frac{\partial}{\partial t}\zeta_1(\omega,t)&=&\hat L_1 \zeta_1(\omega,t) + \sqrt{2}\beta \omega \left[{\cal A}_{10} \zeta_0(\omega,t) 
+ {\cal A}_{12} \zeta_2(\omega,t) \right]
\nonumber \\
\frac{\partial}{\partial t}\zeta_2(\omega,t)&=&\hat L_2 \zeta_2(\omega, t) + \sqrt{2} \beta {\cal A}_{21} \zeta_1(\omega,t) 
\end{eqnarray}
where 
\begin{equation}
\label{eq: 4.3.14}
\hat L_j \equiv \frac{1}{2}\left[\partial_\omega^2 + \frac{2}{\omega} \partial_\omega  
- \frac{j(j+1)}{\omega^2}-\omega^2+3\right]
\ .
\end{equation}

\section{Asymptotic properties of the correlation function}
\label{sec: 4} 

In three dimensions we are only able to determine the spectrum 
of $\hat {\cal H}$ (defined by equation (\ref{eq: 3.3.2})) by 
analytical methods in the limits $\beta\to 0$ and $\beta\to \infty$.
This section considers various asymptotic approximations for the correlation function. 

\subsection{Short-time limit}
\label{sec: 4.0}

The correlation function $C(t)=\langle {\bf n}(t)\cdot{\bf n}(0)\rangle$ can be calculated 
by using the initial condition ${\bf n}(0)={\bf e}_3$ and then computing $\langle \cos\theta\rangle$.
For short times, the polar angle is approximated by $\theta=(\omega_1^2+\omega_2^2)^{1/2}t+O(t^2)$.
In the short-time limit, therefore
\begin{equation}
\label{eq: 4.0.1}
C(t)=\langle 1-\frac{1}{2}\theta^2+\ldots \rangle=1-\frac{1}{2}\langle \omega_1^2+\omega_2^2\rangle t^2 +\ldots
\end{equation}
Using equation (\ref{eq: 1.3}), we have $\langle \omega_i^2\rangle=D/\gamma$. 
The leading order behaviour of the correlation function is therefore
\begin{equation}
\label{eq: 4.0.2}
C(t)=1-\frac{Dt^2}{\gamma}+O(t^3)=1-\beta^2 \bar t^2+O(t^3)
\end{equation}
where $\bar t$ is the dimensionless time defined by (\ref{eq: 3.1.10}).
This result is valid for all $\beta$.

\subsection{Diffusive ($\beta \to 0$) limit}
\label{sec: 4.1}

In the limit $D/\gamma^3\ll 1$, the angle of ${\bf n}$ diffuses with 
diffusion coefficient ${\cal D}=D/\gamma^2$. The solution of the diffusion
equation on the surface of a sphere is expressed in terms of 
spherical harmonics $Y_{lm}(\theta,\phi)$, for which the eigenvalues
of the Laplacian are $-l(l+1)$:
\begin{equation}
\label{eq: 4.1.1}
P(\theta,\phi,t)=\sum_{l=0}^\infty \sum_{m=-l}^l a_{lm}\exp[-l(l+1)Dt/\gamma^2]Y_{lm}(\theta,\phi)
\ .
\end{equation}
The required coefficients are obtained using orthogonality of spherical harmonics: 
$a_{lm}=Y_{l0}({\bf e}_3)\delta_{m0}$, so that $a_{lm}=\delta_{m0}\delta_{l1}\sqrt{3/4\pi}$.
The expectation value of $\cos(\theta)$ therefore has a simple exponential decay:
\begin{equation}
\label{eq: 4.1.2}
\langle \cos \theta \rangle=\sqrt{\frac{4\pi}{3}}\langle Y_{10}(t)\rangle=\exp(-2Dt/\gamma^2)=\exp(-2\beta^2\bar t)
\ .
\end{equation}
This approximation does not have the correct limiting behaviour as $t\to 0$, 
which is given by (\ref{eq: 4.0.2}). The following approximation to the numerically determined correlation function approaches (\ref{eq: 4.1.2}) at $\beta \to 0$ for all $t$ and has the correct quadratic behaviour at $t=0$:
\begin{equation}
\label{eq: 4.1.3}
C(t)\approx \exp\left(-\frac{2\beta^2 \bar t^2}{\sqrt{4+\bar t^2}}\right)
\ .
\end{equation}

\subsection{Large $\beta $ limit}
\label{sec: 4.2}

In the limit $D/\gamma^3\gg 1$, the short-time behaviour of the correlation function
is determined by tumbling motion with fixed angular momemtum. Consider the solution of the
equation of motion $\dot {\bf n}=
\mbox{\boldmath$\omega$}\wedge {\bf n}$ in the case
where $\mbox{\boldmath$\omega$}=\omega {\bf e}_{\omega}$ is constant and 
where the initial direction is ${\bf n}_0={\bf e}_3$. The solution is
\begin{equation}
\label{eq: 4.2.1}
{\bf n}(t)=\mbox{\boldmath$a$}+\mbox{\boldmath$b$}\cos\omega t+\mbox{\boldmath$c$}\sin\omega t
\end{equation}
where 
\begin{equation}
\label{eq: 4.2.2}
\mbox{\boldmath$a$}=({\bf e}_\omega\cdot {\bf n}_0){\bf e}_\omega
\ ,\ \ \ 
\mbox{\boldmath$b$}={\bf n}_0-\mbox{\boldmath$a$}
\ ,\ \ \ 
\mbox{\boldmath$c$}=\frac{\mbox{\boldmath$a$}\wedge \mbox{\boldmath$b$}}
{|\mbox{\boldmath$a$}\wedge \mbox{\boldmath$b$}|}|\mbox{\boldmath$b$}|
\end{equation}
are three mutually orthogonal vectors.
It follows that
\begin{equation}
\label{eq: 4.2.3}
{\bf n}(t)\cdot {\bf n}(0)=|{\bf n}_0\cdot {\bf e}_\omega|^2
+\cos \omega t [1-|{\bf n}_0\cdot{\bf e}_\omega|^2]
\ .
\end{equation}
Now integrate over the distribution of angular momentum to obtain:
\begin{eqnarray}
\label{eq: 4.2.4}
\langle {\bf n}(t)\cdot{\bf n}(0)\rangle&=&\int {\rm d}\mbox{\boldmath$\omega$}\ 
P(\mbox{\boldmath$\omega$})\left[\frac{\omega^2_z}{\omega^2}+\cos\omega t \left(1-\frac{\omega_z^2}{\omega^2}\right)\right]
\nonumber \\
&=&\left(\frac{\gamma}{2\pi D}\right)^{3/2}\int_{-\infty}^\infty {\rm d}\omega_x \int_{-\infty}^\infty 
{\rm d}\omega_y \int_{-\infty}^\infty {\rm d}\omega_z\ \exp(-\gamma \mbox{\boldmath$\omega$}^2/2D)
\left[\frac{\omega^2_z}{\omega^2}+\left(1-\frac{\omega_z^2}{\omega^2}\right)\cos \omega t\right]
\nonumber \\
&=&\left(\frac{\gamma}{2\pi D}\right)^{3/2}\int_0^\infty {\rm d}\omega \int_0^\pi {\rm d}\theta\ 2\pi \omega^2\sin\theta\exp(-\gamma \omega^2/2D)\left[\cos^2\theta+\sin^2\theta\cos\omega t\right]
\nonumber \\
&=&4\pi\left(\frac{\gamma}{2\pi D}\right)^{3/2}\int_0^\infty {\rm d}\omega\ \omega^2
\exp(-\gamma \omega^2/2D)\left[\frac{1}{3}+\frac{2}{3}\cos\omega t\right]
\nonumber \\
&=&\frac{1}{3}+\frac{2}{3}\left(1-\frac{Dt^2}{\gamma}\right)\exp(-Dt^2/2\gamma)
\ .
\end{eqnarray}
This shows that in the three-dimensional case, when $\beta =\sqrt{D/\gamma^2}\gg 1$ the correlation
function decays to $\langle {\bf n}(t)\cdot{\bf n}(0)\rangle =1/3$ on a rapid timescale, 
$\tau_1=(\beta\gamma)^{-1}$, due to motions with different frequencies getting out of phase. 

On longer timescales the value of $\mbox{\boldmath$\omega$}$ 
fluctuates, and the correlation function then decays to zero on a slower timescale describing the decay of 
correlations of $\mbox{\boldmath$\omega$}$. A precise understanding this limit 
requires us to carry out a more sophisticated analysis, as will be done
in section \ref{sec: 4.4} below.  
Before addressing this issue we consider 
how ${\bf n}(t)$ behaves when $\mbox{\boldmath$\omega$}$ varies
slowly. We show that the angle between ${\bf n}$ and $\mbox{\boldmath$\omega$}$ is an adiabatic
invariant of the dynamics of (\ref{eq: 3.1.1}). This shows that the decay of correlations of ${\bf n}(t)$
is governed by the diffusion of the direction of $\mbox{\boldmath$\omega$}$, implying that
the timescale for the decay of correlations of ${\bf n}(t)$ is $O(\gamma^{-1})$ in the
limit as $\beta\to\infty$.

To show that the angle $\theta$ between ${\bf n}(t)$ and $\mbox{\boldmath$\omega$}$ is an adiabatic
invariant, consider the time evolution of
\begin{equation}
\label{eq: 4.2.5}
f\equiv {\bf n}\cdot \mbox{\boldmath$\omega$}=\omega \cos\theta \equiv \omega z
\end{equation}
where $\omega=|\mbox{\boldmath$\omega$}|$ and where the second equality defines $z=\cos\theta$. 
From (\ref{eq: 3.1.1}), the time derivative of $f$ is
\begin{equation}
\label{eq: 4.2.6}
\frac{{\rm d}f}{{\rm d}t}={\bf n}\cdot \frac{{\rm d}\mbox{\boldmath$\omega$}}{{\rm d}t}
=\left[\frac{z}{\omega}\mbox{\boldmath$\omega$}+\delta {\bf n}(t)\right]\cdot \frac{{\rm d}\mbox{\boldmath$\omega$}}{{\rm d}t}
\end{equation}
where $\delta {\bf n}(t)$ oscillatates on a timescale $\omega^{-1}$ about a mean value which is equal to zero
when $\mbox{\boldmath$\omega$}$ is constant. When evaluating the drift of $f$, we neglect this rapidly
oscillating term, and write 
\begin{equation}
\label{eq: 4.2.7}
\left\langle\frac{{\rm d}f}{{\rm d}t}\right\rangle\approx \frac{z}{\omega} \mbox{\boldmath$\omega$}\cdot \frac{{\rm d}\mbox{\boldmath$\omega$}
}{{\rm d}t}
=\frac{z}{2\omega}\frac{{\rm d}\omega^2}{{\rm d}t}
\ .
\end{equation}
Alternatively, from the definition $f=z\omega$, we find
\begin{equation}
\label{eq: 4.2.8}
\frac{{\rm d}f}{{\rm d}t}=\omega\frac{{\rm d}z}{{\rm d}t}+z\frac{{\rm d}\omega}{{\rm d}t}=\omega \frac{{\rm d}z}{{\rm d}t}+\frac{z}{2\omega}
\frac{{\rm d}\omega^2}{{\rm d}t}
\ .
\end{equation}
Comparing (\ref{eq: 4.2.7}) and (\ref{eq: 4.2.8}) we see that $\dot z=0$, implying that $\theta $ is an adiabatic invariant,
as stated above.

We have argued that in the limit as $\beta \to \infty$ the direction of ${\bf n}(t)$ is determined
by the evolution of $\mbox{\boldmath$\omega$}$. This indicates that we can determine the 
rate of decay of $C(t)$ by determining the rate of decay of correlations of the angular momentum. The direction vector of the angular momentum, 
${\bf e}_\omega$, exhibits 
diffusion on the surface of the unit sphere with a diffusion 
coefficient ${\cal D}_\omega$ which we determine shortly. In a diffusiuon process, the expectation value of the spherical harmonic $\cos\theta$ decays exponentially: $\langle \cos\theta\rangle\sim \exp(-2{\cal D}_\omega t)$. We might therefore expect that
when $\beta=\sqrt{D/\gamma^3}\gg 1$, the correlation function is well approximated
by
\begin{equation}
\label{eq: 4.2.9}
\langle {\bf n}(t)\cdot {\bf n}(0)\rangle\sim \frac{1}{3}\exp(-\lambda t)
+\frac{2}{3}\left(1-\frac{Dt^2}{\gamma}\right)\exp(-Dt^2/2\gamma)
\ .
\end{equation}
and that $\lambda=2{\cal D}_\omega$. This 
argument is, however, not satisfactory. When the deviation from
equilibrium of the distribution of ${\bf n}(t)$
is well approximated by the spherical harmonic $\cos \theta$, the corresponding 
distribution of ${\bf e}_\omega$ might be a quite
different combination of spherical harmonics. This 
question is most effectively addressed by the application of 
group theory to the Fokker-Planck equation, using results from sections
\ref{sec: 3.2} and \ref{sec: 4.3} above.

We conclude this section by determining the diffusion coefficient for diffusion of the 
direction of $\mbox{\boldmath$\omega$}$. This direction is the unit vector 
${\bf e}_\omega=\mbox{\boldmath$\omega$}/\omega$, which diffuses on the unit 
sphere with diffusion coefficient ${\cal D}_\omega$. The change in the 
${\bf e}_\omega$ in a short time $\delta t$ is 
\begin{equation}
\label{eq: 4.2.10}
\delta {\bf e}_\omega=\frac{1}{\omega^2}
\left[\omega \delta \mbox{\boldmath$\omega$}-(\mbox{\boldmath$\omega$}\cdot \delta \mbox{\boldmath$\omega$})\mbox{\boldmath$\omega$}\right]
\end{equation}
so that the expectation value of the square of the rotation angle is 
\begin{equation}
\label{eq: 4.2.11}
\langle \delta {\bf e}_\omega^2\rangle=\frac{2}{\omega^2}
\langle \delta \omega^2_i \rangle 
=\frac{4D\delta t}{\omega^2}
\ .
\end{equation}
The required diffusion coefficient is obtained by averaging this over the known distribution of $\mbox{\boldmath$\omega$}$: 
\begin{eqnarray}
\label{eq: 4.2.12}
{\cal D}_\omega&=&\frac{\langle \delta {\bf e}_\omega^2\rangle}{4\delta t}
=D\int {\rm d}\mbox{\boldmath$\omega$}\ \frac{P(\mbox{\boldmath$\omega$})}{\omega^2}
\nonumber \\
&=&D\left(\frac{\gamma}{2\pi D}\right)^{3/2}\int_0^\infty {\rm d}\omega\ 4\pi \omega^2 \frac{\exp(-\gamma\omega^2/2D)}{\omega^2}
=\gamma
\ .
\end{eqnarray}
The correlation function for the direction of the angular momentum therefore
deays as 
\begin{equation}
\label{eq: 4.2.13}
\langle {\bf e}_\omega (t)\cdot {\bf e}_\omega (0)\rangle=\exp(-2{\cal D}_\omega t)=\exp(-2\gamma t)
\ .
\end{equation}
Numerical evidence indicates that the correlation function 
$C(t)$ decays at a different rate when $\beta \gg 1$: the decay rate
in (\ref{eq: 4.2.9}) is found to be $\lambda\approx 1.56\gamma$, rather
than $\lambda=2\gamma$. The difference is explained in the following
section, \ref{sec: 4.4}.

\subsection{Asymptotic solution of slowest mode}
\label{sec: 4.4}

Here we use results derived from symmetry considerations 
(in sections \ref{sec: 3.2} and \ref{sec: 4.3}) to identify the
slowest decaying modes in the limit as $\beta\to\infty$.
The objective is to determine solutions of (\ref{eq: 4.3.13}) which decay exponentially in time, so that $\zeta_j(\omega,t)=\exp(-\lambda t) ~ a_j(\omega)$.
The functions $a_j(\omega)$ satisfy the eigenvalue equation

\begin{eqnarray}
\label{eq: 4.4.1}
\hat L_0\, a_0(\omega) + \sqrt{2} \beta \omega  {\cal A}_{01} a_1(\omega) &=&\lambda\ a_0(\omega)
\nonumber \\
\hat L_1\, a_1(\omega) + \sqrt{2} \beta \omega \left[{\cal A}_{10} a_0(\omega) 
+ {\cal A}_{12} a_2(\omega) \right]&=&\lambda\ a_1(\omega)
\nonumber \\
\hat L_2\, a_2(\omega) + \sqrt{2} \beta \omega {\cal A}_{21} a_1(\omega)
&=& \lambda\ a_2(\omega) 
\end{eqnarray}
The structure of the operator (\ref{eq: 4.3.14}) implies that
\begin{equation}
\label{eq: 4.4.2}
\lim_{\omega\to 0}\frac{a_j(\omega)}{\omega^j}=C_j
\end{equation}
for some constants $C_j$.

We are interested in the most slowly decaying solutions, which requires determining the
eigenvalue $\lambda$ with the largest real part. The structure of (\ref{eq: 4.4.1}) suggests that
the decay rates are expected to increase in proportion to $\beta $ as $\beta\to \infty$. However, our discussion 
in section \ref{sec: 4.2} indicates that there should be eigenfunctions which have a slow deacy rate, $\lambda=O(\gamma)$ as $\beta\to\infty$. In order to identify these slow modes, note that the matrix
${\cal A}_{ij}$ has a null eigenvector: this matrix is
\begin{equation}
\label{eq: 4.4.3}
\{{\cal A}_{ij}\}=\left(\begin{array}{ccc}
0 & \sqrt{\frac{2}{3}} & 0\cr 
-\sqrt{\frac{2}{3}} & 0 & \sqrt{\frac{1}{3}}\cr 
0 & -\sqrt{\frac{1}{3}} & 0 
\end{array}\right)
\end{equation}
which has a null vector, $(1,0,2)$. The only way to obtain an eigenfunction
with an eigenvalue which remains bounded as $\beta\to\infty$ is to
assume that throughout most of the range of $\omega$, we have 
\begin{equation}
\label{eq: 4.4.4}
{\cal A}_{10}\,a_0(\omega)+{\cal A}_{12}\,a_2(\omega)\approx 0
\ .
\end{equation}
Using this approximation, the equation for $a_2(\omega)$ can be written
\begin{equation}
\label{eq: 4.4.5}
\left[\hat L_2+\rho \hat L_0\right]a_2=(1+\rho)\,\lambda a_2
\end{equation}
where 
\begin{equation}
\label{eq: 4.4.6}
\rho=\frac{{\cal A}_{21}{\cal A}_{12}}{{\cal A}_{01}{\cal A}_{10}}
=\frac{1}{2}
\ .
\end{equation}
However, we have
\begin{equation}
\label{eq: 4.4.7}
\hat L_j=\hat L_0+\frac{j(j+1)}{\omega^2}
\end{equation}
so that $a_2(\omega)$ is an eigenfunction of
\begin{equation}
\label{eq: 4.4.8}
\hat L_j=\hat L_0+\frac{6}{1+\rho}\frac{1}{\omega^2}
\end{equation}
with eigenvalue $\lambda$. This is an operator of the form (\ref{eq: 4.4.7})
with an angular momentum quantum number $j$ which satisfies 
$j(j+1)=6/(1+\rho)=4$, that is,
\begin{equation}
\label{eq: 4.4.10}
j=\frac{\sqrt{17}-1}{2}
\ .
\end{equation}
Thus it is argued that, in the limit as $\beta \to \infty$, there exist
modes for which the eigenvalues are $O(\gamma)$, which satisfy
a radial equation with an {\sl irrational} value of the angular momentum,
given by (\ref{eq: 4.4.10}). It remains to identify the eigenvalues
associated with this equation. The first step is to factor out the boundary  
condition at $\omega = 0$, writing  
\begin{equation}
\label{eq: 4.4.11}
a_j(\omega) = \omega^j \Phi(\omega)
\ .
\end{equation}
Then it is useful to remove the Gaussian factor from the solution, writing $\Phi(\omega) = \exp(-\omega^2/2) \phi(\omega^2)$. The equation for $\phi(u)$ (with $u = \omega^2$) is:
\begin{equation}
\label{eq: 4.4.12} 
u \phi'' + (j + 3/2 - u) \phi'  + \frac{(\lambda - j)}{2}\phi = 0
\ .
\end{equation}
Well behaved polynomial solutions exist only when $(\lambda - j) = 2n$  
($n$ integer). Inserting the factor of $\gamma$ which is required 
when we return to dimensioned equations, we conclude that the  
eigenvlaue which gives the slowest rate of decay is
\begin{equation}
\label{eq: 4.4.13}
\lambda=-\gamma j=-\frac{\gamma(\sqrt{17}-1)}{2}
\end{equation}
and the full set of eigenvalues of the problem defined by
equation (\ref{eq: 4.4.12}) is
\begin{equation}
\label{eq: 4.4.14}
\lambda_{n}=-\gamma(j+2n)\ ,\ \ \ \ n=0,1,2,\ldots
\ .
\end{equation}
The eigenfunctions can also be expressed in terms of generalised Laguerre
polynomials, so that the Fokker-Planck equation may be regarded as exactly 
solvable in the limit as $\beta\to\infty$. 

We have imposed two apparently incompatible conditions on the solutions
$a_j(\omega)$, namely (\ref{eq: 4.4.2}) and (\ref{eq: 4.4.4}). We conclude this 
section by considering how these are reconciled.  
Consider the nature of the solutions $a_j(\omega)$ close to $\omega=0$.
Dimensionally, the problem is analogous to one with the following structure:
${\rm d}^2 \Phi/{\rm d}\omega^2 + \beta \omega \Phi = 0$.
The derivative terms becomes dominant when $\omega < \beta^{-1/3}$, so that
the approximation (\ref{eq: 4.4.4}) fails and (\ref{eq: 4.4.2}) becomes applicable
when $\omega \beta^{1/3}\ll 1$. 
The problem can be treated by applying standard asymptotic expansion 
methods \cite{BendOrsz+99}. Alternatively, one can determine the solutions of 
the problem by expanding the solution on a conveniently complete basis, and by
truncation, reduce the solution of equations (\ref{eq: 4.4.1}) to a matrix equation, 
as explained in Appendix A.

\begin{figure}
\centering
\begin{tabular}{cc}
\epsfig{file=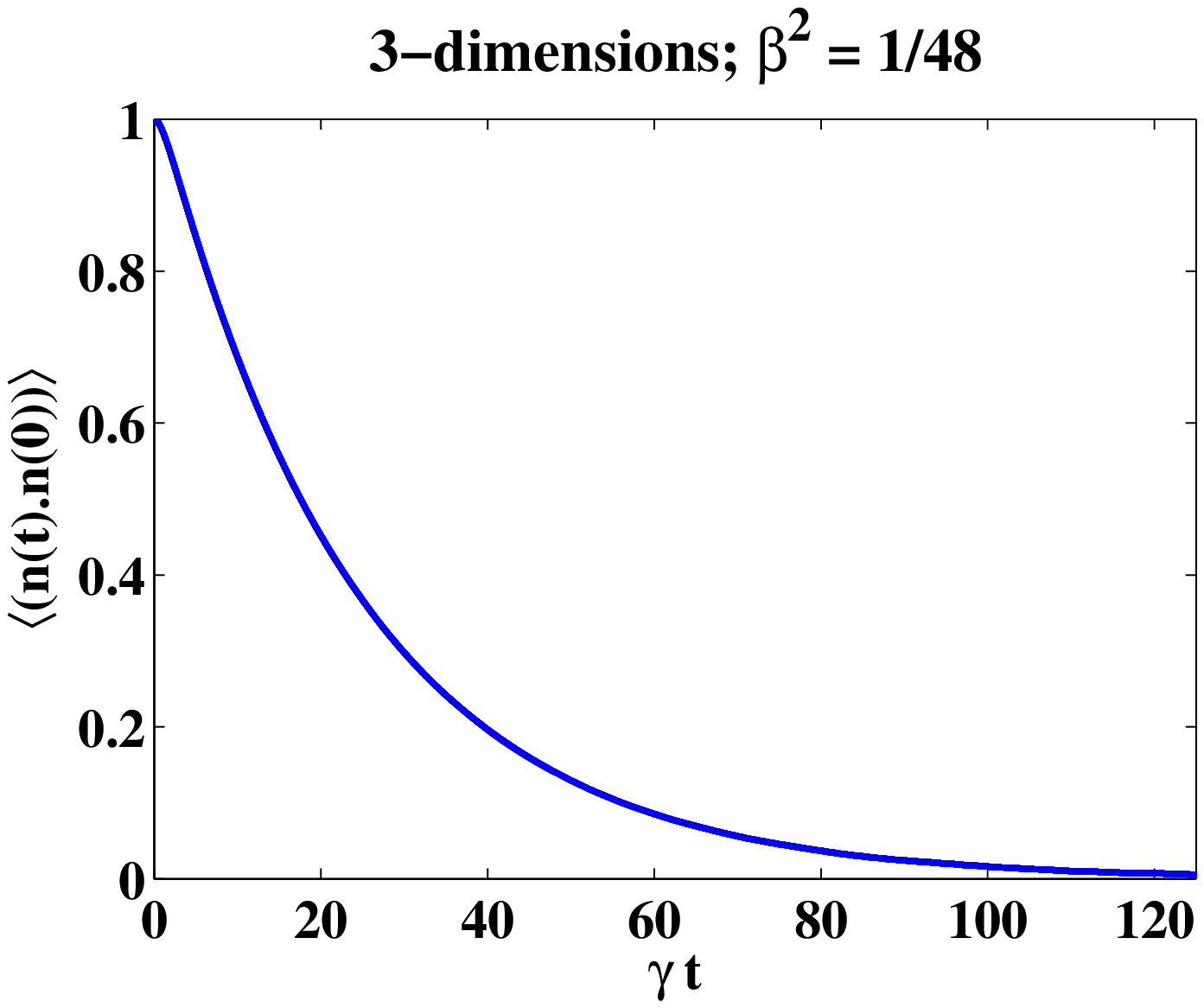,width=0.4\linewidth,clip=} & 
\epsfig{file=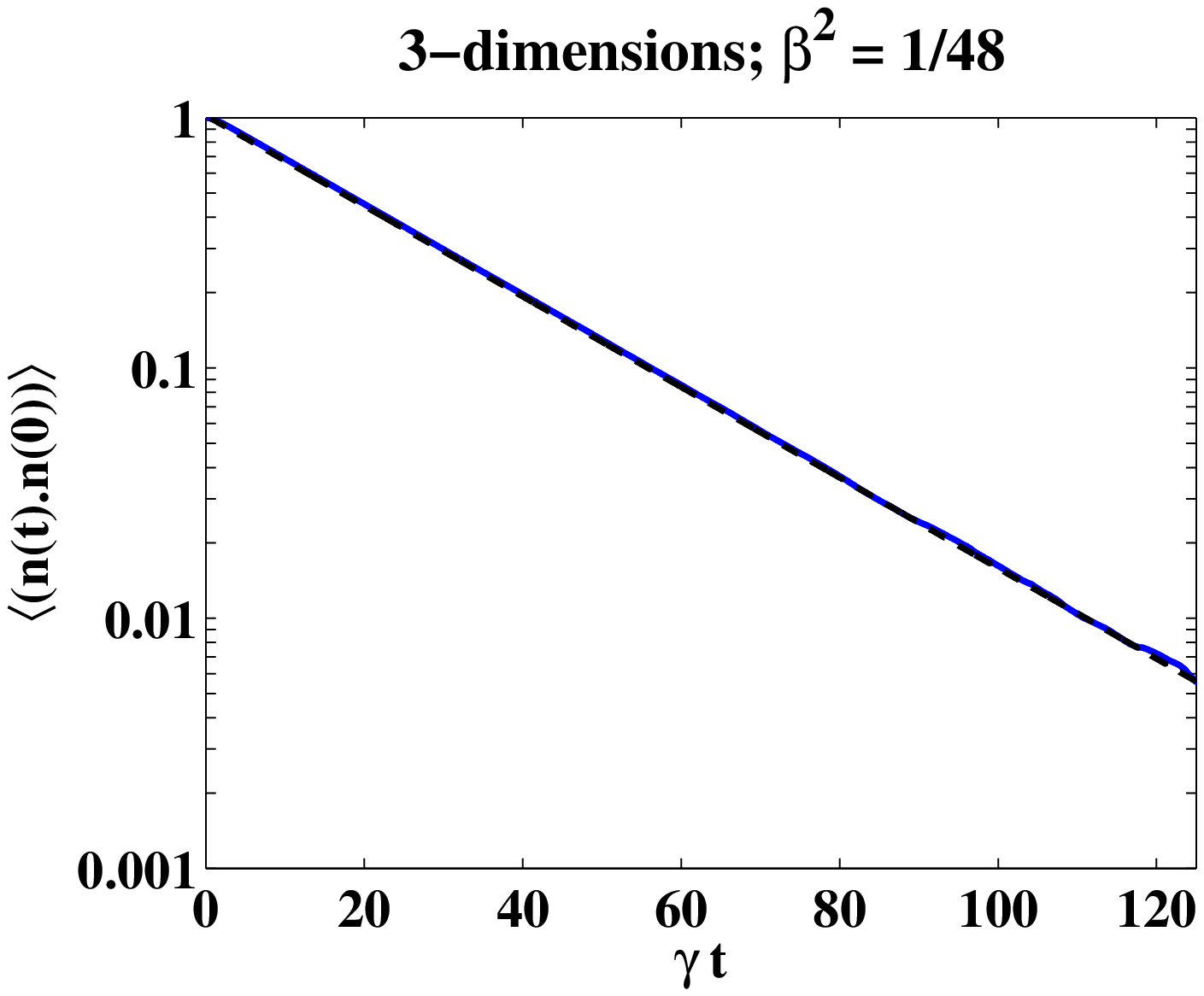,width=0.4\linewidth,clip=} \\
\epsfig{file=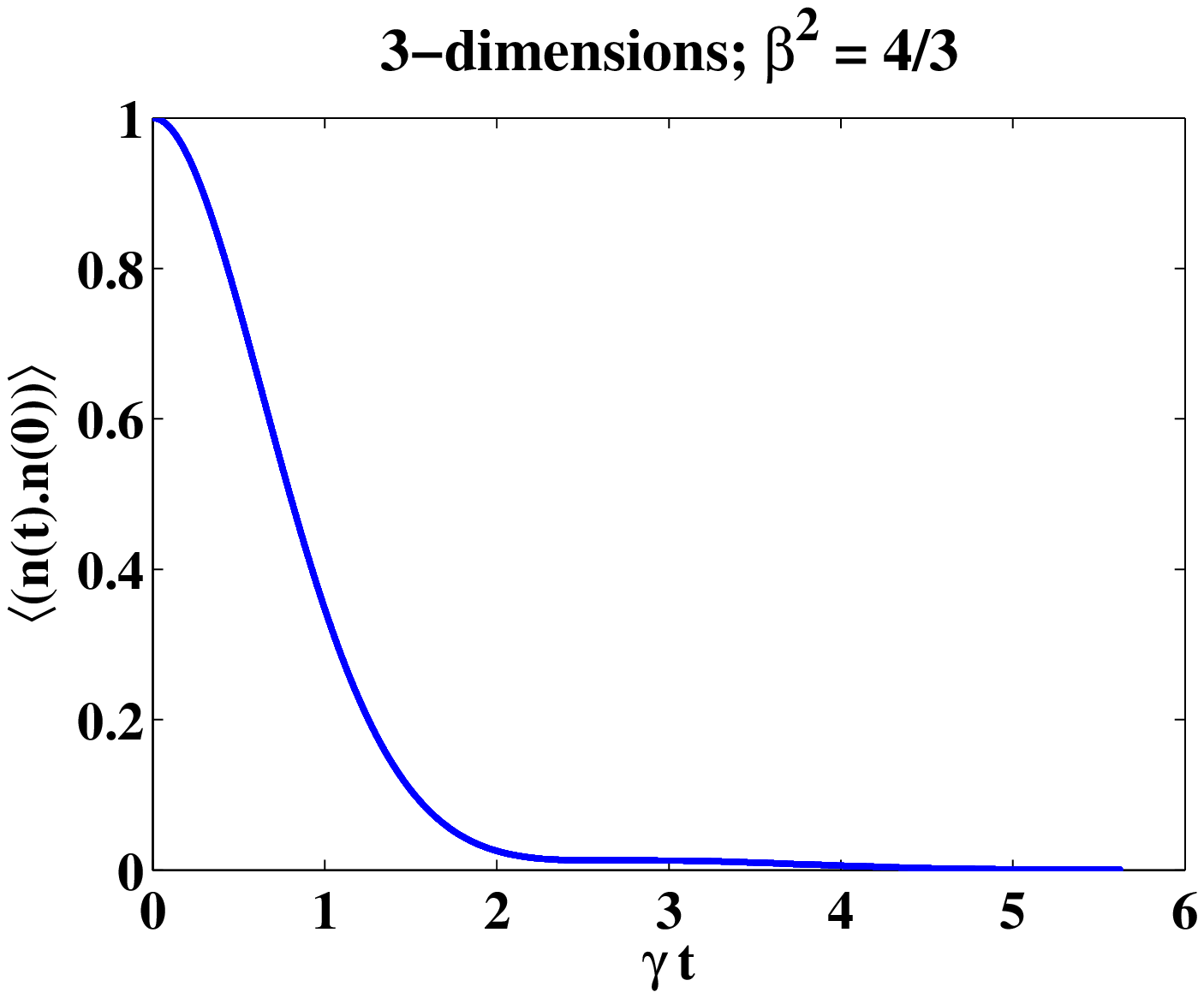,width=0.4\linewidth,clip=} &
\epsfig{file=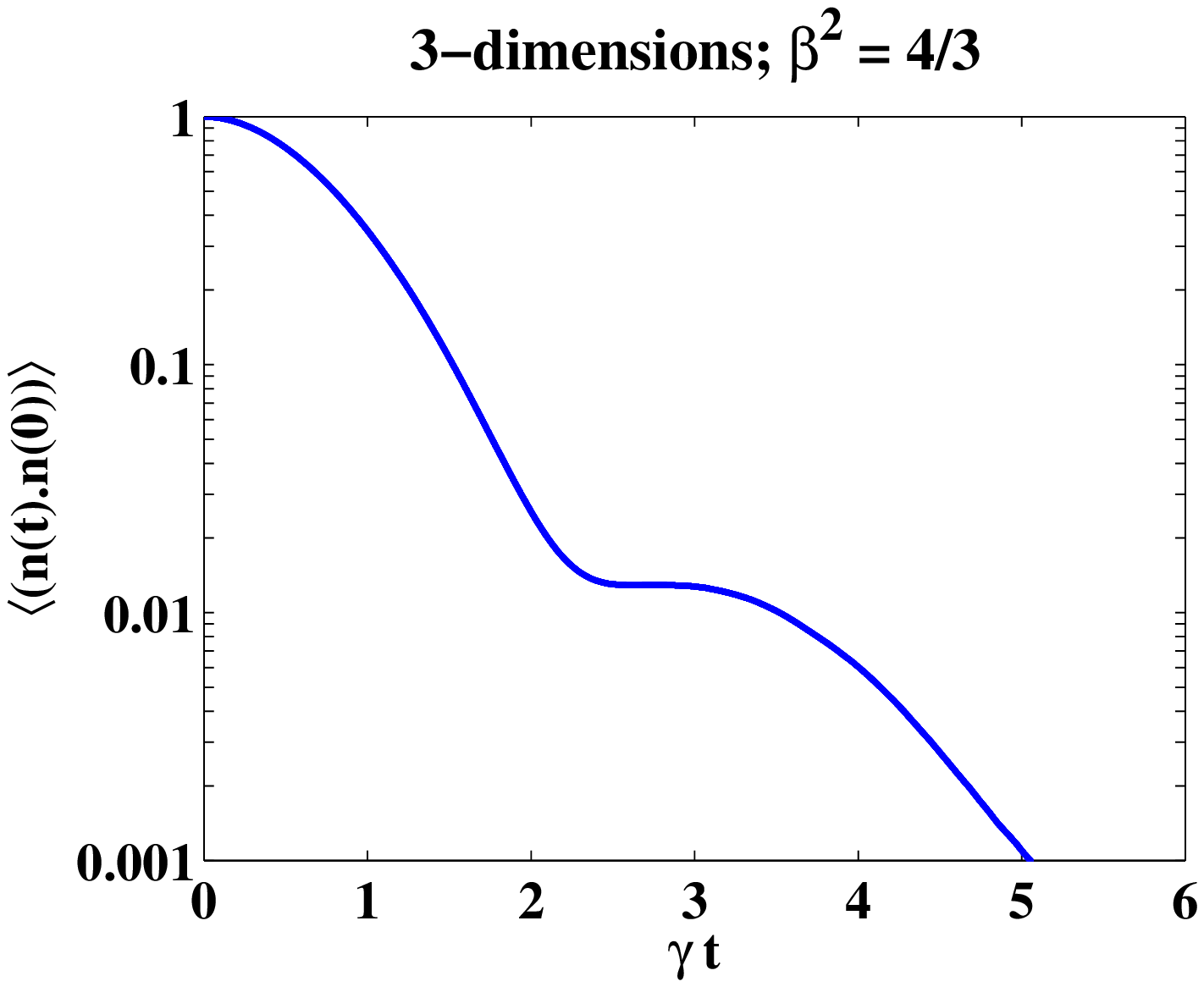,width=0.4\linewidth,clip=}\\
\epsfig{file=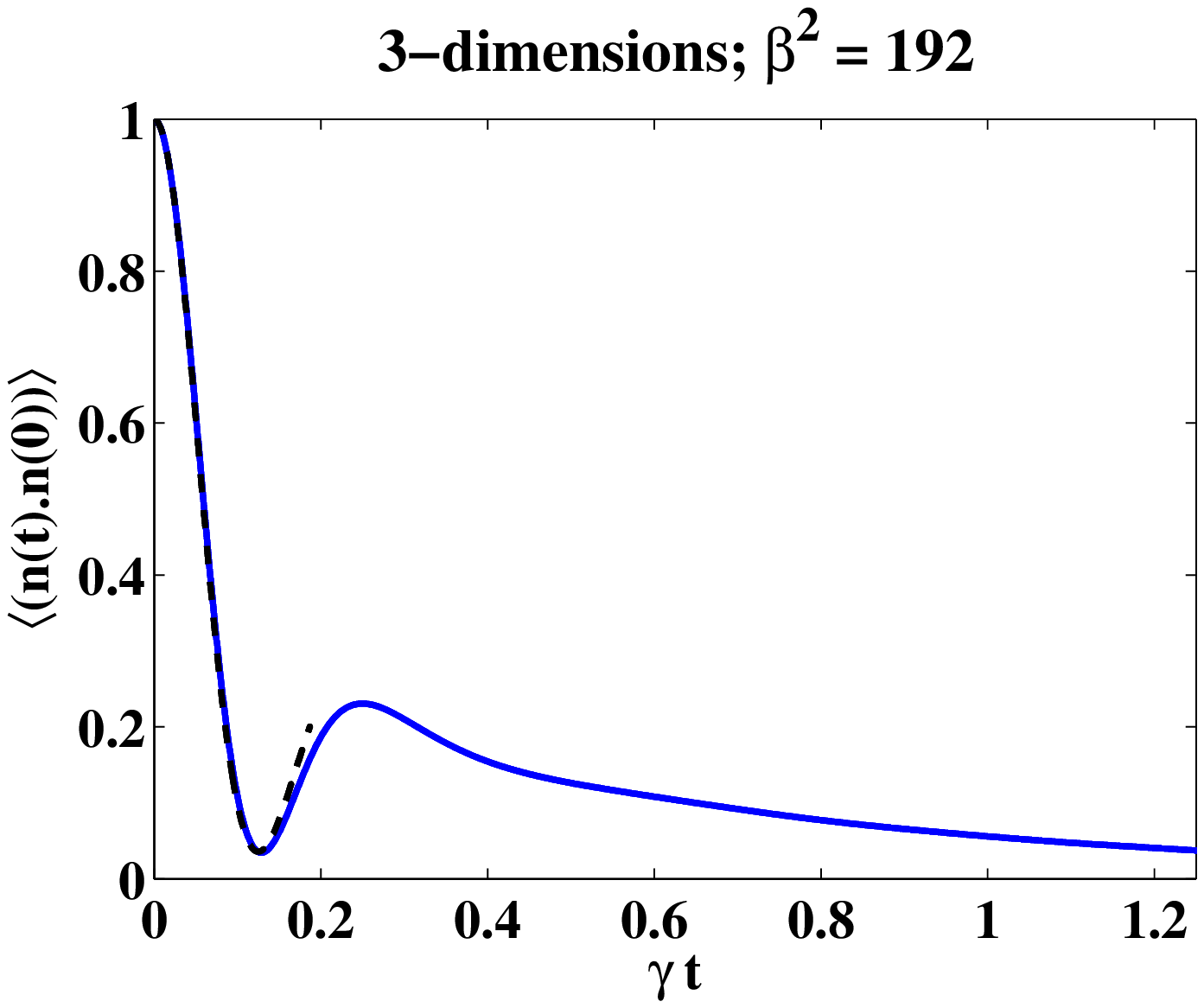,width=0.4\linewidth,clip=} &
\epsfig{file=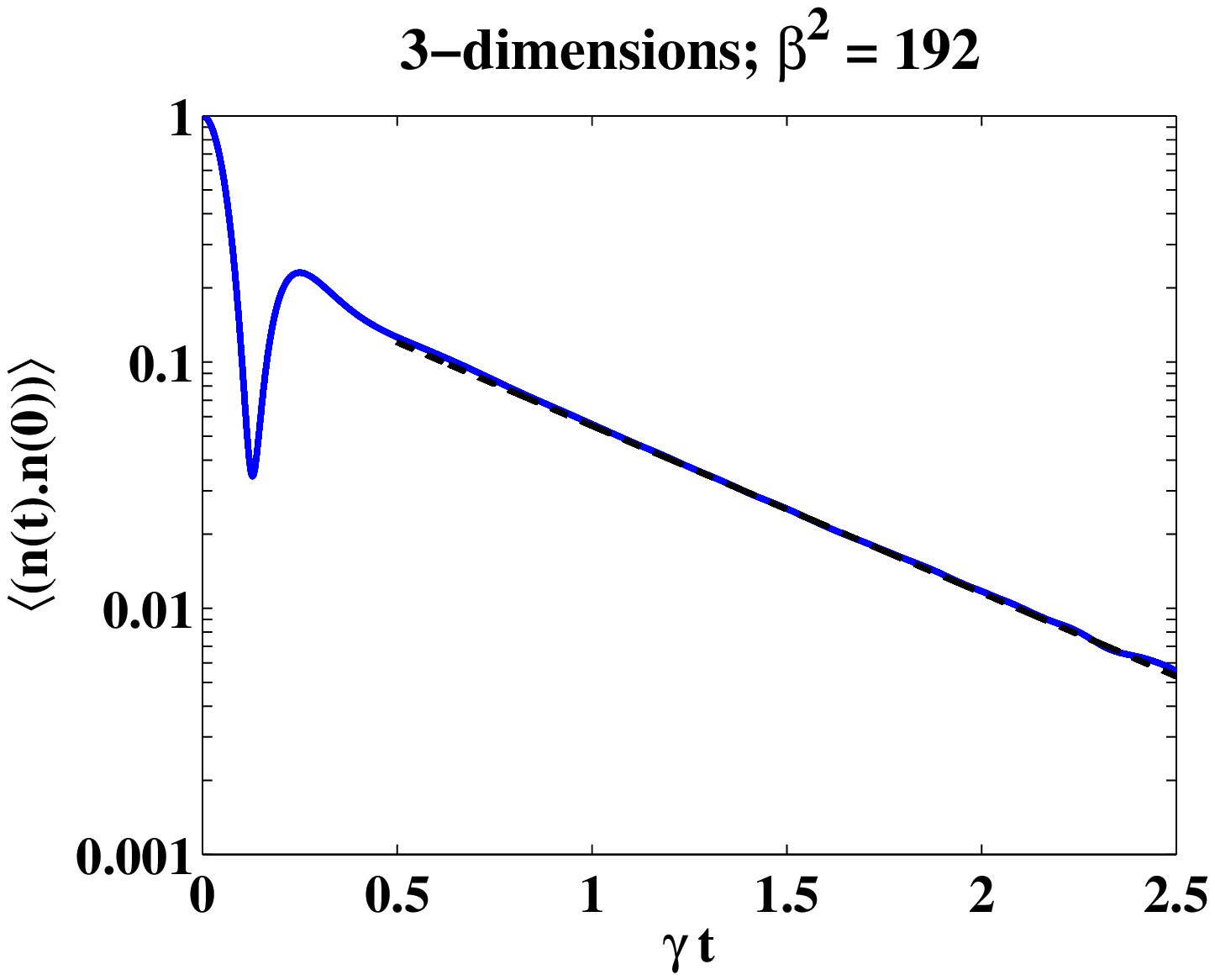,width=0.4\linewidth,clip=}
\end{tabular}
\caption{\label{fig: 2} Correlation function for the spherical Ornstein-Uhlenbeck
process, for three values of $\beta^2=D/\gamma^3$: 
$\beta^2 = 1/48$ (upper row), $\beta^2 = 4/3$ (middle row) and 
$\beta^2 = 192$ (lower row).
The data for $\beta^2=1/48$ show good agreement with the exponential approximation to the correlation 
function, equation (\ref{eq: 4.1.2}), which is applicable in the limit as 
$\beta \to 0$. The linear plot for $\beta^2=192$ shows good agreement with the transient 
described by equation (\ref{eq: 4.2.9}), and the logarithmic plot 
demonstrates that the correlation function $C(t)$ behaves at long
times as $C(t) \propto \exp( - (\sqrt{17}-1) \gamma t/2)$, as 
predicted by equation (\ref{eq: 4.4.13}).
}
\end{figure}

\begin{figure}
\centering
\begin{tabular}{ccc}
\epsfig{file=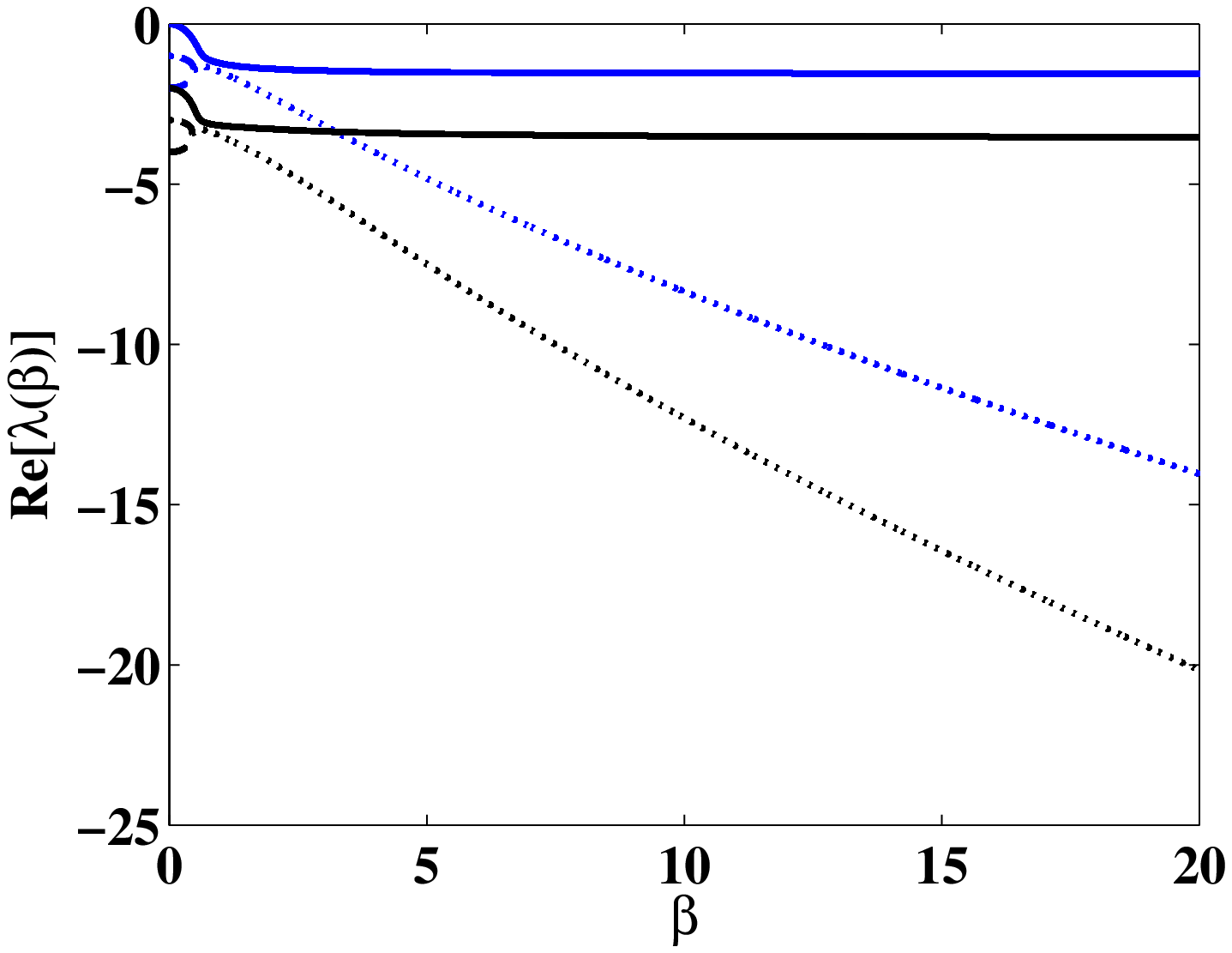,width=0.3\linewidth,clip=} & 
\epsfig{file=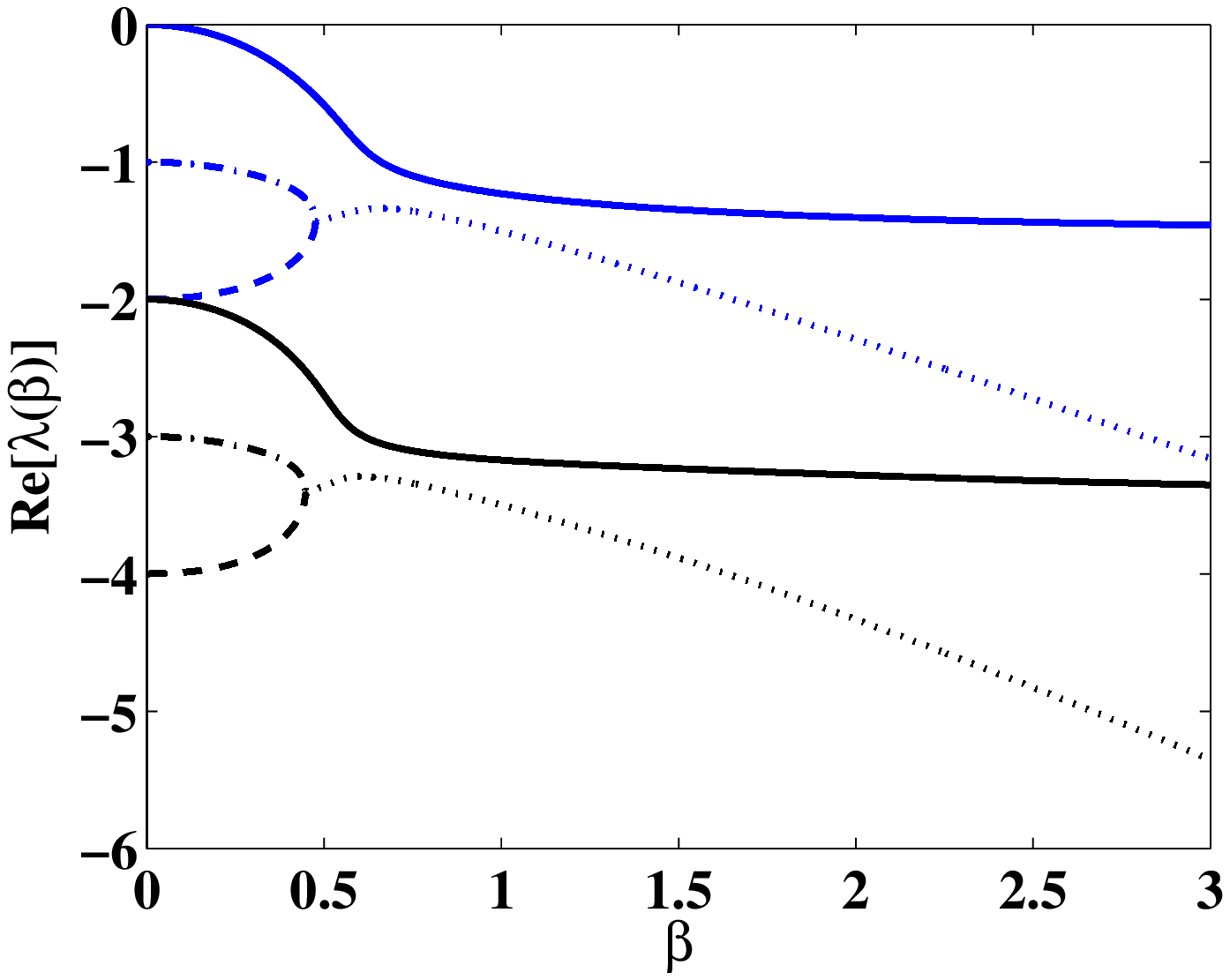,width=0.3\linewidth,clip=}&
\epsfig{file=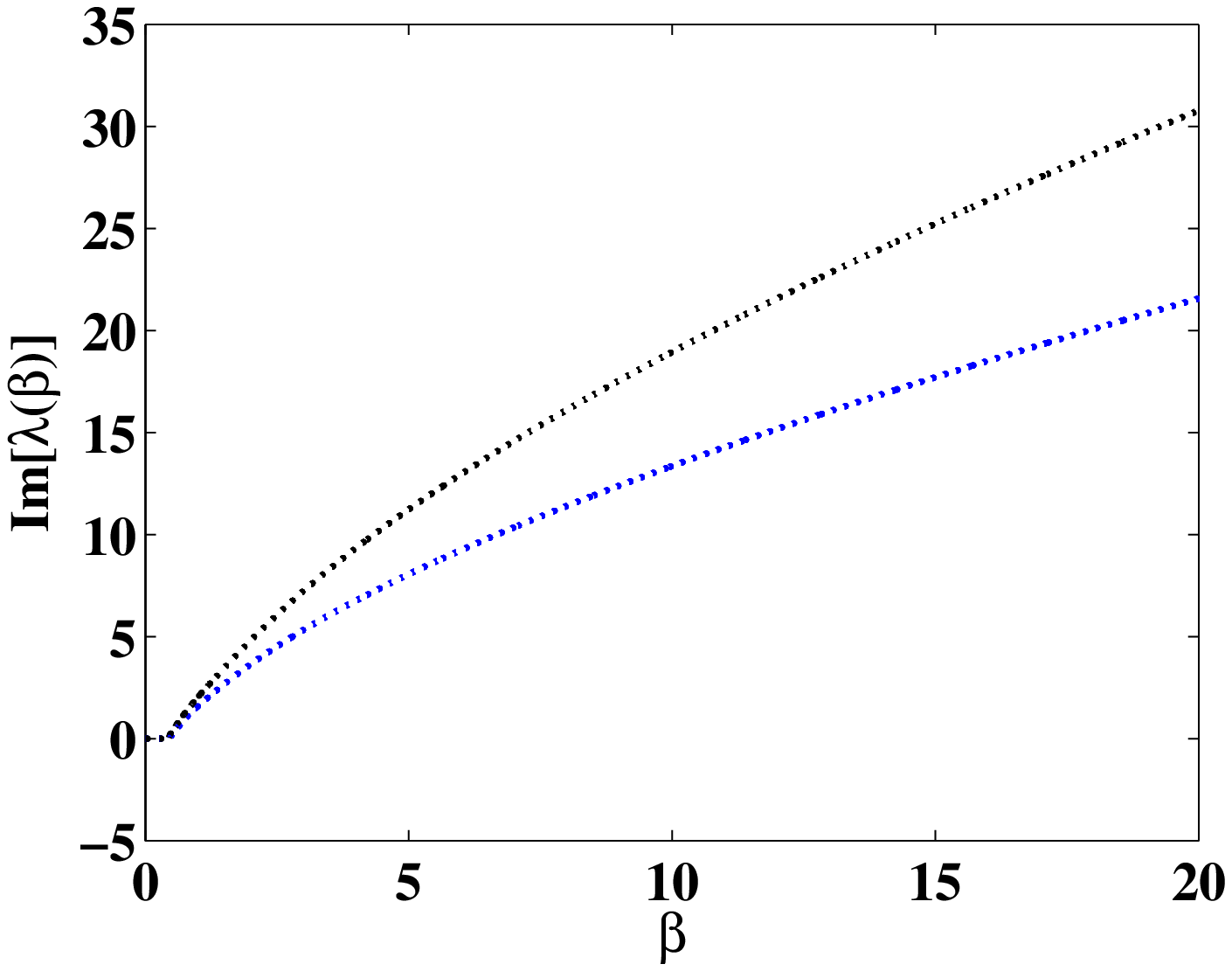,width=0.3\linewidth,clip=}
\end{tabular}
\caption{\label{fig: 3} The six slowest decaying eigenvales from the spectrum of the 
Fokker-Planck operator of the spherical Ornstein-Uhlenbeck process as a function of $\beta$.
Left: real part of the eigenvlaues, showing the separation into levels which diverge
as $\beta\to\infty$ and levels which approach a finite limit. Centre: real parts at greater 
magnification. At very small values of $\beta$, all eigenvalues are real. 
Pairs of branches collide at a finite value of $\beta$, giving rise to pairs of 
complex congugates eigenvectors, whose real parts are shown by the dashed lines.
Right: positive imaginary parts of the eigenvalues.} 
\end{figure}

\begin{figure}
\centering
\begin{tabular}{cc}
\epsfig{file=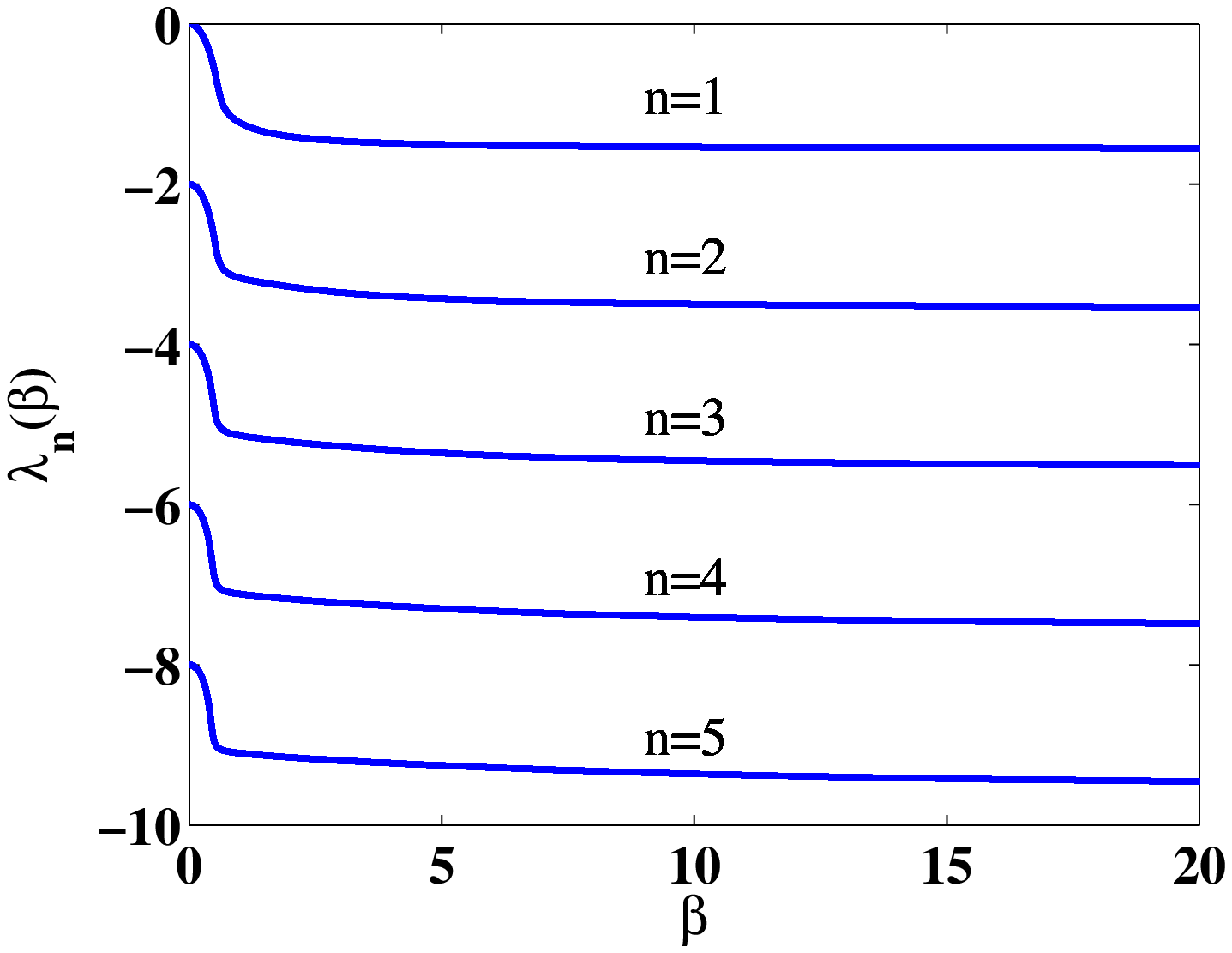,width=0.4\linewidth,clip=} & 
\epsfig{file=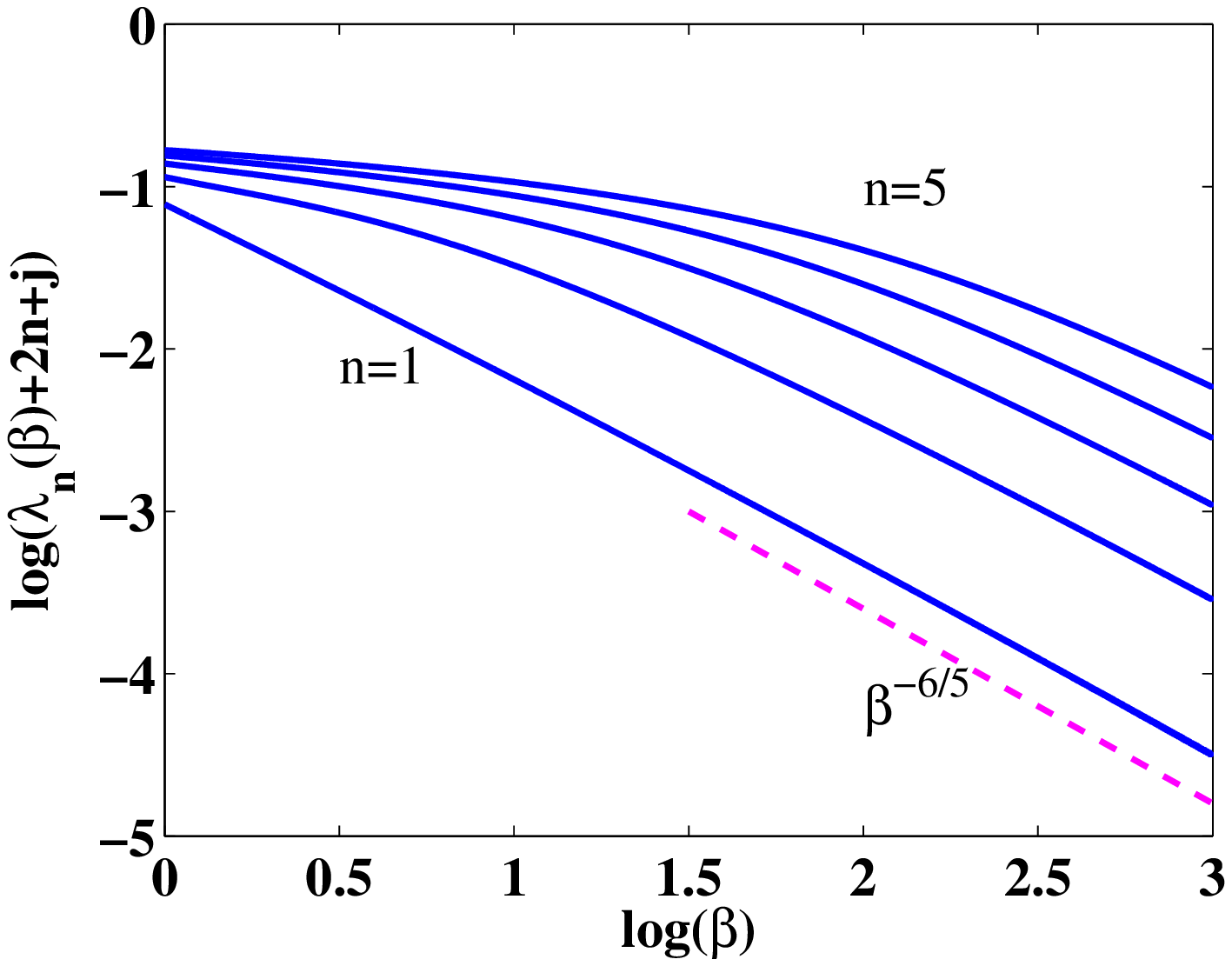,width=0.4\linewidth,clip=}
\end{tabular}
\caption{\label{fig: 4} The real-valued eigenvalues approach the spectrum $\lambda_n=j+2n$ in the
limit as $\beta\to \infty$, in accord with equation (\ref{eq: 4.4.13}). The right-hand panel shows
the convergence towards this limit as $\beta \rightarrow \infty$.}
\end{figure}

Figure \ref{fig: 2} shows the numerically
computed correlation functions for the three-dimensional
Ornstein-Uhlenbeck process, compared with various asymptotic approximations. 
We also investigated the spectrum of the Fokker-Planck operator
numerically. Using the eigenfunctions of the spherical harmonic oscillator
as a basis set, this operator can be represented by an infinite-dimensional
matrix.  The formulae for the matrix elements are given in appendix A. 
We find that the spectrum of finite dimensional truncations converge
as the size of the basis set increases, and we identify the converged
eigenvalues with elements of the spectrum of the Fokker-Planck operator.
Figures \ref{fig: 3} and \ref{fig: 4} illustrate the dependence of the eigenvalues 
upon $\beta$. 

\section{Applications to random tumbling}
\label{sec: 5}

This paper has described a model for the statistics of a unit vector ${\bf n}(t)$ moving randomly 
but smoothly over the surface of a sphere. The model is a generalisation of the Ornstein-Uhlenbeck 
process to a spherical geometry. It has the merit of being susceptible to analytical treatments,
including an exact solution in two dimensions.
It is of interest to consider the possible applications of this model, and the extent 
to which the model provides an accurate description of various systems.  

The spherical Ornstein-Uhlenbeck model could be used to describe the 
tumbling of an object in a turbulent fluid flow. Examples include 
the rotational motion of rocks or dust grains in turbulent circumstellar discs \cite{Gut+10},  
or of ice crystals in a convecting atmosphere \cite{Pru+97}. The rotational motion of such 
bodies can influence their growth by aggregation or their evaporation by 
exposure to a source of radiant heat. Recently it has become possible to make 
detailed experimental studies of the orientation of a neutrally buoyant sphere 
in a turbulent fluid by matching images of an irregularly painted ball to 
photographs taken with the ball in a well defined orientation \cite{Zim+11,Zim+11b}. 
The direction ${\bf n}(t)$ of one axis through the sphere can be followed as a function of 
time. The model could also be used for the fluctuations of direction vector  
${\bf n}(t)$ of a rod-like body in a turbulent fluid. 

These remarks raise the question as to whether the spherical Ornstein-Uhlenbeck
model will give an accurate description of the tumbling motion of a body. In the case of
a small body in a turbulent flow, we argue below that the statistical properties of the
velocity gradients of turbulence appear to make this a very good model.
The orientation of a small body in a turbulent flow with velocity field 
$\mbox{\boldmath$u$}(\mbox{\boldmath$r$}(t),t)$ responds to the gradients of the the 
velocity field, evaluated along the trajectory $\mbox{\boldmath$r$}(t)$ of the body (which can be
assumed to be advected with the fluid). The velocity gradients form a matrix ${\bf A}(t)$, with 
elements $A_{ij}(t)=\partial u_i/\partial r_j(\mbox{\boldmath$r$}(t),t)$. It is convenient
to write ${\bf A}=\mbox{\boldmath$\Omega$}+{\bf S}$, where $\mbox{\boldmath$\Omega$}$, the vorticity
tensor, is antisymmetric and where ${\bf S}$, the strain-rate, is symmetric. The equation of motion 
for the direction vector ${\bf n}(t)$ of a microscopic ellipsoidal object in a fluid flow was obtained
by Jeffery \cite{Jef22}. It can be written in the form
\begin{equation}
\label{eq: 5.1}
\frac{{\rm d}{\bf n}}{{\rm d}t}=\mbox{\boldmath$\Omega$}(t){\bf n}+\frac{\alpha^2-1}{\alpha^2+1}
\left[{\bf S}(t){\bf n}-({\bf n}\cdot {\bf S}(t){\bf n}){\bf n}\right]
\end{equation}
where $\alpha$ is the axis ratio of the ellipsoid. The same equation of motion applies to general
axisymmetric bodies, provided they are small compared to any characteristic lengthscale 
of the flow \cite{Bre62}, but the relation between $\alpha$ and the shape of the object
is not known in general. In the case of a spherical particle or other object with $\alpha=1$, the 
equation of motion (\ref{eq: 5.1}) is of the same form as equation (\ref{eq: 3.1.1}), if we interpret
the vorticity $\mbox{\boldmath$\omega$}(t)$ as being the antisymmetric tensor corresponding
to the axial vector $\mbox{\boldmath$\omega$}$, with elements related by 
$\omega_i=\epsilon_{ijk}\Omega_{jk}$. 

Furthermore, the Lagrangian correlation function of the 
vorticity in turbulent flows has been investigated using simulations 
of turbulent flows by several authors \cite{Gir+90,Bru+98,Pum+11}. The elements have mean value equal
to zero and appear to be statistically independent. It was found that the correlation
function of each elements can be fitted quite accurately by an exponential function:
\begin{equation}
\label{eq: 5.2}
\langle \Omega_{ij}(t)\Omega_{ij}(t')\rangle =\frac{D_{\rm v}}{\gamma_{\rm v}}\exp(-\gamma_{\rm v} |t-t'|)
\ .
\end{equation}
The decay rate $\gamma_{\rm v}$ is of the order of the inverse of the Kolmogorov 
timescale $\tau_{\rm K}$ of the turbulent flow, which is the shortest timescale of the fluid motion:
$\tau_{\rm K}=\sqrt{\nu/{\cal E}}$, where $\nu$ is the kinematic viscosity and where
${\cal E}$ is the rate of dissipation per unit mass. Numerical evidence indicates that
$\gamma_{\rm v}\approx 1/(8.5\tau_{\rm K})$ at large Reynolds numbers \cite{Pum+11}.
The diffusion coefficient $D_{\rm v}$ can be related to $\gamma_{\rm v}$ by various
kinematic constraints (discussed in \cite{Pum+11}), giving 
$D_{\rm v}=\gamma_{\rm v}/12\tau_{\rm K}^2$.
These considerations suggest that the spherical Ornstein-Uhlenbeck model should
describe tumbling of a small object in a turbulent flow, with a \lq universal' value for the 
persistence angle 
\begin{equation}
\label{eq: 5.3}
\beta_{\rm turb}=\sqrt{\frac{D_{\rm v}}{\gamma_{\rm v}^3}}=\frac{1}{12 \gamma^2_{\rm v}\tau_{\rm K}}
\approx 2.4 
\ .
\end{equation}
However, we cannot conclude that correlation function $C(t)=\langle {\bf n}(t)\cdot{\bf n}(0)\rangle$ 
for objects in a turbulent fluid will correspond to that of our spherical 
Ornstein-Uhlenbeck model, because vorticity may have very different
temporal variation, compared to the Ornstein-Uhlenbeck model, but still
have the same correlation function.
This point is illustrated by a calculation in appendix B, where we analyse 
a model for random motion on a circle, in which the angular velocity is determined by a telegraph 
process (examples of this type of model are considered in \cite{Sha+78,Fal+07}). The telegraph
model has a correlation function of angular velocity which is an exponential function, equivalent
to that of the circular Ornstein-Uhlenbeck process. However we show that the correlation function 
$\langle {\bf n}(t)\cdot{\bf n}(0)\rangle$ is very different for the two models. 
We conclude that the extent to which the spherical Ornstein-Uhlenbeck process is a good
description of tumbling in a turbulent fluid must be tested by numerical simulations 
of turbulence. Our own numerical investigations on the tumbling of microscoipic
particles in turbulence indicate that the spherical Ornstein-Uhlenbeck model is not
a good model for their correlation function \cite{Pum+11}. This conclusion is consistent with studies
by Shin and Koch \cite{Shi+05}, who presented data for $\langle {\bf n}(t)\cdot {\bf n}(0)\rangle$ in 
simulations of rod-like objects in fully developed driven turbulent flows. 

It is only in cases where the statistics of the angular momentum fluctuations are 
a precise match to the spherical Ornstein-Uhlenbeck process that reliable predictions 
can be made about the correlation function defined by (\ref{eq: 1.6}). In the case of
an object tumbling in a very dilute gas, such as a small rock in the circumstellar
disc of the star, the spherical Ornstein-Uhlenbeck model 
may be a very good description of the evolution
of the angular momemtum. Bombardment by microscopic dust grains can provide random
impulses which change the angular momentum in the same way as the white-noise fluctuations
in (\ref{eq: 3.1.2}). And the damping due to motion in an extremely dilute gas is 
proportional to the relative velocity \cite{Eps24}, consistent with the linear damping term in 
equation (\ref{eq: 3.1.2}). We conclude that a small body tumbling in a very dilute gas 
is one example where the spherical Ornstein Uhlenbeck model is an excellent description
of a physical process.

\section{Concluding remarks}
\label{sec: 6}

Our study was motivated by recent works, which have characterized the 
orientation of particles transported by a turbulent flow 
\cite{Zim+11,Zim+11b,Shi+05,Par+11,Bez+10}. 
These processes are naturally described by the random motion on a spherical surface.
The work here has focused on arguably the simplest model for smooth random 
motion on a sphere: the direction $\bf{n}$ rotates with an angular velocity
$\mbox{\boldmath$\omega$}$, which evolves according to an Ornstein-Uhlenbeck process. 

The solution of this model depends on 
only one dimensionless parameter, which we refer to here as $\beta$, the
persistence angle, which characterizes the rotation occuring during one correlation time of $\mbox{\boldmath$\omega$}$.
We have characterised evolution of ${\bf n}(t)$ by analysing its correlation function  
$\langle {\bf n}(0) {\bf n}(t) \rangle$.

In two dimensions, we have obtained an explicit expression for the 
correlation function, in terms of very elementary functions. This was 
achieved by completely diagonalizing the Fokker-Planck operator and hence 
writing the correlation function as a series, whose sum can be
explicitly determined. 

In contrast, the motion on the three-dimensional sphere is more involved. This
is largely due to the more complicated structure of the rotation group in 
three dimensions. Quite generally, the components of ${\bf n}$ perpendicular
to $\mbox{\boldmath$\omega$}$ rapidly rotate, hence decorrelate, whereas the component parallel
to $\mbox{\boldmath$\omega$}$ remains unchanged, at least when 
$\mbox{\boldmath$\omega$}$ is constant. In the 
limit when the persistance angle is large, this leads to a two time-scale
dynamics: a fast decorrelation of ${\bf n}$ is 
observed, corresponding to the components of ${\bf n}$ perpendicular
to $\mbox{\boldmath$\omega$}$, followed by a much slower decorrelation of ${\bf n}$, corresponding
to the component parallel to $\mbox{\boldmath$\omega$}$. Whereas the fast decorrelation can be 
understood quantitatively by using elementary considerations, the description
of the decorrelation of ${\bf n}$ at large times requires a determination of the
largest eigenvalue of the Fokker-Planck operator. We have computed here 
the eigenvalues relevant to the long term evolution of 
$\langle  \bf{n}(0) \cdot {\bf n}(t) \rangle$. Interestingly, in the large
$\beta $ limit, the problem reduces to a quantum harmonic oscillator with
a irrational angular momentum.

The Ornstein-Uhlenbeck model is closely related to the equation describing 
the orientational degrees of freedom of small particles in turbulent flows \cite{Jef22,Shi+05},
and numerical studies show that the vorticity of turbulent flows also has an exponential correlation
\cite{Bru+98,Pum+11}. However, we have observed here that caution should observed 
in applying our model to rotation by turbulence. In two dimensions we showed 
that the correlation function of $\bf n$ takes a very different form when the angular 
velocity is generated by a telegraph process, despite the fact that the correlation 
function of $\omega$ are identical to our model.

In summary, the notion of the \lq persistence angle' introduced 
here appears to be the most relevant parameter characterising random rotation.
The Ornstein-Uhlenbeck model is the simplest description of random motion 
on a sphere, and it will surely find signoficant applications, beyond the example
considered at the end of section \ref{sec: 5}. However, it is a poor model 
for rotations of small bodies driven by hydrodynamic turbulence \cite{Pum+11}.

{\sl Acknowledgements}. MW thanks the ENS Lyon for a for 
visiting position. AP was supported by the  
french Agence Nationale pour la Recherche under contract DSPET, and by  
IDRIS for computer ressources.

\section{Appendix A: Matrix representation}
\label{sec: app}

\subsection{Decomposition and projection.}
\label{subsec: appA}

The aim of this appendix is to project the system of partial
differential equations (\ref{eq: 4.3.13}) on the complete set of eigenfunctions
of the spherical harmonic oscillator operator. 
The operators $\hat L_l$, introduced in equation (\ref{eq: 4.3.14}) 
correspond to the radial part of the equation for the
harmonic oscillator operators, equations (\ref{eq: 3.3.harm_osc}, \ref{eq: 3.3.4})
associated with angular momentum quantum number $l = 0$, $1$ and $2$ 
respectively.

The eigenvalues of $\hat L_l$ are thus $-(2n + l)$, $n$ being 
the quantum number 
characterizing energy, and the corresponding radial eigenfunction being:
\begin{equation}
\phi_n^{(l)}(r) = {\cal N}_{nl} ~ \exp(-r^2/2) ~ r^l ~ L_n^{(l+1/2)} (r^2)
\label{def_phi}
\end{equation}
where $L_n^{(\alpha)}$ is the generalized Laguerre polynomial \cite{Lan+58}, 
and the normalization constant is given by equation (\ref{eq: 4.4.1}).
The following discussion uses properties of the generalised Laguerre polynomials 
which are discussed in \cite{Wolfram}.

For each value of $l$, these eigenstates are orthogonal to each other, in 
the sense that:
\begin{eqnarray}
( \phi_n^{(l)} , \phi_m^{(l)}) & = & \delta_{nm} 
\ .
\label{ortho_int}
\end{eqnarray}

One can therefore expand the functions $\psi^{l)}i$ in series of the $\phi^{(l)}_n$:
\begin{equation}
\psi^{(l)}(r) = \sum_{n=0}^\infty a_n^{(l)} ~ \phi_n^{(l)} (r) 
\ .
\label{expansion}
\end{equation}
This can then be inserted into the set of equations (\ref{eq: 4.4.1}). Then,
the equation corresponding to angular momentum $l$ is projected on the set
of modes $\phi_n^{(l)}(r)$. This leads to:
\begin{eqnarray}
\label{prod_scal_eq1}
(\phi_n^{(0)}, \hat L_0 \psi^{(0)}) - 2/\sqrt{3} \beta ( \phi_n^{(0)}, r \psi^{(1)}) &=& \lambda
(\phi_n^{(0)}, \psi^{(0)}) 
\nonumber \\
(\phi_n^{(1)}, \hat L_1 \psi^{(1)})  +  2/\sqrt{3} \beta ( \phi_n^{(1)}, r \psi^{(0)}) 
- \sqrt{2/3} \beta (\phi_n^{(1)}, r \psi^{(2)}) &=& \lambda
(\phi_n^{(1)}, \psi^{(1)}) 
\nonumber \\
(\phi_n^{(2)}, \hat L_2 \psi^{(2)}) + \sqrt{2/3} \beta ( \phi_n^{(2)}, r \psi^{(1)}) &=& \lambda
(\phi_n^{(2)}, \psi^{(2)}) 
\end{eqnarray}
Clearly, 
\begin{equation}
( \phi_n^{(l)} , \hat L_l \psi^{(l)})  = - a_n^{(l)} ~ ( 2 n + l) 
\ .
\label{prod_scal_simpl}
\end{equation}
The scalar products of $( \phi_n^{(0)} , r \psi^{(1)})$ and $( \phi_n^{(1)} , r \psi^{(0)} )$,
on one hand, and
$( \phi_n^{(1)} , r \psi^{(2)} ) $ and $( \phi_n^{(2)} , r \psi^{(1)} )$ on the other hand
involve the calculations of the two integrals:
\begin{eqnarray}
I_{n,m}^{0,1} & \equiv & \int_0^{\infty} \exp(-r^2) r^{0 + 1} 
\times  r \times r^2 
L_n^{(1/2)}(r^2) L_m^{(3/2)}(r^2)\ {\rm d}r 
\nonumber \\
& = & \frac{1}{2} \int_0^{\infty} \exp(-u) u^{3/2} L_n^{(1/2)}(u) L_m^{(3/2)}(u)\ {\rm d}u
\label{prod_scal1}
\end{eqnarray}
and:
\begin{eqnarray}
I_{n,m}^{1,2} & \equiv & \int_0^{\infty} \exp(-r^2) r^{1 + 2} 
\times  r \times  r^2 
L_n^{(3/2)}(r^2) L_m^{(5/2)}(r^2)\ {\rm d}r 
\nonumber \\
& = &  \frac{1}{2} \int_0^{\infty} \exp(-u) u^{5/2} L_n^{(3/2)}(u) L_m^{(5/2)}(u)\ {\rm d}u
\ .
\label{prod_scal2}
\end{eqnarray}
These two integrals can be easily computed, by using the following
relation between generalized Laguerre polynomials:
\begin{equation}
L_n^{(\alpha)}(x) = L_n^{(\alpha + 1)} (x) - L_{n-1}^{(\alpha + 1)}(x)
\label{recurs}
\end{equation}
and the normalization integral (\ref{eq: 3.3.8}).
One thus finds: 
\begin{eqnarray}
I_{n,m}^{0,1}
& = & \frac{1}{2} \int_0^{\infty} \exp(-u) u^{3/2} L_m^{(3/2)}(u) 
(L_n^{(3/2)}(u) - L_{n-1}^{(3/2)}(u))\ {\rm d}u 
\nonumber \\
& = & ( \delta_{n,m} - \delta_{n-1,m} ) \frac{\Gamma(5/2 + m)}{ 2 ~ m!} 
\label{ps_1}
\end{eqnarray}
and
\begin{eqnarray}
I_{n,m}^{1,2}
& = & \frac{1}{2} \int_0^{\infty} \exp(-u) u^{5/2} L_m^{(5/2)}(u) 
(L_n^{(5/2)}(u) - L_{n-1}^{(5/2)}(u))\ {\rm d}u 
\nonumber \\
& = & ( \delta_{n,m} - \delta_{n-1,m} ) \frac{\Gamma(7/2 + m)}{ 2 ~ m!} 
\ .
\label{ps_2}
\end{eqnarray}
Using these results, the set of equations (\ref{prod_scal_eq1})
reduces to a matrix equation, with a relatively simple (band-) structure, as 
we explain below.

\subsection{Matrix equations}
\label{subsec: appB}

It is now a simple matter to rewrite the matrix equations for the quantities
$a_n^{(l)}$, defined in equation (\ref{expansion}).
Specifically, from Eq. (\ref{prod_scal_eq1}), one obtains the system of equations:
\begin{eqnarray}
-2 n a_n^{(0)} - 2/\sqrt{3} \beta
\sum_{m = 0}^{\infty} A(n,m) a_m^{(1)} &=&
\lambda a_n^{(0)}
\nonumber \\
\label{matrix_eqn_0}
-( 2 n + 1) a_n^{(1)} 
+ 2\sqrt{3} \beta\sum_{m = 0}^{\infty} B(n,m) ~ a_m^{(0)} 
- \sqrt{2/3} \beta  \sum_{m = 0}^{\infty} C(n,m) ~ a_m^{(2)} 
&=& \lambda a_n^{(1)}
\label{matrix_eqn_1}
\nonumber \\
-( 2 n + 2) a_n^{(2)} 
+ \sqrt{2/3} \beta \sum_{m = 0}^{\infty} D(n,m)  a_m^{(1)} 
&=& \lambda a_n^{(2)}
\label{matrix_eqn_2}
\end{eqnarray}
where 
\begin{eqnarray}
A(n,m) = \frac{I_{n,m}^{0,1}}{{\cal N}_{n0} {\cal N}_{m1}}&\ \ \ \ &
B(n,m) = \frac{I_{m,n}^{0,1}}{{\cal N}_{m0} {\cal N}_{n1} }
\nonumber \\
C(n,m) = \frac{I_{n,m}^{1,2}}{{\cal N}_{n1} {\cal N}_{m2} }&\ \ \ \ &
D(n,m) = \frac{I_{m,n}^{1,2}}{{\cal N}_{m1} {\cal N}_{n2} } 
\ .
\label{def_D}
\end{eqnarray}
In fact, in view of the structure of the scalar products 
Eq. (\ref{ps_1},\ref{ps_2}), the
matrix equations Eq. (\ref{matrix_eqn_0})
contain in fact very few terms. 
Explicitly,
\begin{eqnarray}
\label{eq: matrix-eqn}
-2 n a_n^{(0)} -   2\beta/\sqrt{3} \Bigl[
\Bigl( A(n,n) ~ a_n^{(1)} - A(n,n-1) ~ a_{n-1}^{(1)} \Bigr) \Bigr]& =&
\lambda a_n^{(0)}
\nonumber \\
-(2 n+1) a_n^{(1)} + \beta/\sqrt{3}\Bigl[ 
2\Bigl( B(n,n) ~ a_n^{(0)} - B(n,n+1) ~ a_{n+1}^{(0)} \Bigr) &&
\nonumber \\
-\Bigl( C(n,n) ~ a_n^{(2)} - C(n,n-1) a_{n-1}^{(2)} \Bigr) \Bigr]
&=& \lambda a_n^{(1)}
\nonumber \\
-(2 n+2) a_n^{(2)} + \sqrt{2/3}\beta \Bigl[ 
\Bigl( D(n,n) ~ a_n^{(1)} - D(n,n+1) ~ a_{n+1}^{(1)} \Bigr) \Bigr] 
&=& \lambda a_n^{(2)}
\ .
\end{eqnarray}
The structure of the matrix defined by equations (\ref{eq: matrix-eqn}), 
is easy to program in a routine that can diagonalize a real matrix, with the infinite-dimensional matrix truncated to a finite
size by including only coefficients $a_n^{(l)}$ with $n\le n_{\rm max}$. Although the structure of the 
matrix is very sparse -- the matrix a simple band structure -- a general routine from NAG has been used 
to determine the spectrum. We checked that the eigenvalues converge as the number of coefficients
$n_{\rm max}$ increases. We were able to obtain the eigenvalues with the largest real parts, {\i.e.}, the one that
correspond to the slowest decay of the correlation function. 
 
\section{Appendix B: Telegraph-noise model}
\label{sec: appB}

When solving physical problems it is often tacitly assumed that the correlation 
function of a stochastic signal is sufficient to characterise its properties.
For example, if we were to use a different stochastic process to generate 
the angluar velocity, we might expect that the correlation function 
$C(t)=\langle {\bf n}(t)\cdot{\bf n}(0)\rangle$ would be little
changed if the correlation function of the angular velocity reamins the
same, namely $\langle \omega(t)\omega(0)\rangle =(D/\gamma)\exp(-\gamma|t|)$.
This assumption may not be valid in general. To illustrate this point, 
here we determine the correlation function (\ref{eq: 1.6}) for random 
motion on a circle in the case where the angular 
velocity is generated by a telegraph noise process, which has precisely the same exponential correlation 
function as the Ornstein-Uhlenbeck process. The correlation function of the direction vector,
defined by (\ref{eq: 1.6}), is found to be very different from (\ref{eq: 2.2.7}).

In the telegraph noise model the angular velocity $\omega(t)$ takes just two discrete values,
$\pm \omega_0$ (where $\omega_0$ is a constant). The angular momentum makes random
transitions between these values, with a rate constant $R$ (so that the probability of transition 
in a short time interval of length $\delta t$ is $\delta P=R\delta t$). The rotation angle $\theta $
satisfies $\dot \theta=\omega(t)$, as before. Define $P_\pm (\theta,t)$ to be the probability density
to be located at $\theta $ at time $t$, with $\omega(t)=\pm \omega_0$. The probability 
densities satisfy
\begin{equation}
\label{eq: B.1}
\frac{\partial P_\pm}{\partial t}=-\pm \omega_0\frac{\partial P_{\pm}}{\partial \theta}+R\,P_{\mp}-R\,P_\pm
\ .
\end{equation}
Formally, this equation can be written $\partial_t |P)=\hat {\cal F}\,|P)$, where $|P)$ is a function vector 
representing $(P_+(\theta,t),P_-(\theta,t))$. This equation can be solved by seeking eigenfunctions 
of $\hat {\cal F}$ in the form $|\psi)=\exp({\rm i}n\theta)(c_+,c_-)$. The vector $\mbox{\boldmath$c$}=(c_+,c_-)$ 
is an eigenvector of the $2\times 2$ matrix
\begin{equation}
\label{eq: B.2}
{\bf F}=\left(\begin{array}{cc}
-{\rm i}n\omega_0-R & R \cr
R & {\rm i}n\omega_0-R
\end{array}\right)
\ .
\end{equation}
This matrix has eigenvalues 
\begin{equation}
\label{eq: B.3}
\lambda_{n\pm}=-R \pm \sqrt{R^2-n^2 \omega_0^2}
\ .
\end{equation}
The general solution of (\ref{eq: B.1}) can be expressed as a linear combination
of eigenfunctions:
\begin{eqnarray}
\label{eq: B.4}
\left(\begin{array}{c}
P_+(\theta,t)\cr 
P_-(\theta,t)
\end{array}\right)
&=&\sum_{n=-\infty}^\infty
a_{n+}\exp(\lambda_{n+}t+{\rm i}n\theta)
\left(\begin{array}{c}
R \cr
{\rm i}n\omega_0+\sqrt{R^2-n^2\omega^2_0}
\end{array}\right)
\nonumber \\
&+&
a_{n-}\exp(\lambda_{n-}t+{\rm i}n\theta)
\left(\begin{array}{c}
R\cr
{\rm i}n\omega_0-\sqrt{R^2-n^2\omega_0^2}
\end{array}\right)
\ .
\end{eqnarray}
In order to evaluate the correlation function (\ref{eq: 1.6}) we must
compute $C(t)=\langle \cos\theta \rangle$, with the initial condition
$P_\pm (\theta)=\delta (\theta)/2$. Thus
\begin{eqnarray}
\label{eq: B.5}
C(t)&=&\int_0^{2\pi}{\rm d}\theta\ \cos\theta\,[P_+(\theta,t)+P_-(\theta,t)]
\\
&=&2\pi {\rm Re}\left[a_{1+}\left(R+{\rm i}\omega_0+\sqrt{R^2-\omega_0^2}\right)\exp(\lambda_{1+}t)
+a_{1-}\left(R+{\rm i}\omega_0-\sqrt{R^2-\omega_0^2}\right)\exp(\lambda_{1-}t)\right]
\nonumber
\ .
\end{eqnarray}
The coefficients $a_{n\pm}$ are easily determined from the initial 
distribution:
\begin{equation}
\label{eq: B.5a}
a_{n\pm}=\frac{\sqrt{R^2-n^2\omega_0^2}\pm (R-{\rm i}n\omega_0)}{8\pi R\sqrt{R^2-n^2\omega_0^2}}
\end{equation}
and hence
\begin{equation}
\label{eq: B.6}
C(t)=\frac{1}{2}[\exp(\lambda_+t)+\exp(\lambda_-t)]+\frac{R}{2\sqrt{R^2-\omega_0^2}}[\exp(\lambda_+t)-\exp(\lambda_-t)]
\end{equation} %
where $\lambda_\pm=\lambda_{1\pm}=-R\pm \sqrt{R^2-\omega_0^2}$.
A similar and somewhat simpler calculation gives the correlation function of $\omega(t)$:
\begin{equation}
\label{eq: B.7}
\langle \omega(t)\omega(0)\rangle=\omega_0^2\exp(-2R|t|)
\ .
\end{equation}
Because this correlation function has the same structure as that of the Ornstein-Uhlenbeck
process, we can define the parameters $\omega_0$, $R$ of the telegraph noise model in terms
of the parameters $\gamma$, $D$ of the Ornstein-Uhlenbeck process: by comparison 
of (\ref{eq: B.7})  with (\ref{eq: 1.3}) we have
\begin{equation}
\label{eq: B.8}
\gamma=2R
\ ,\ \ \ \ 
D=2R\omega_0^2
\ ,\ \ \ \ 
\beta=\frac{\omega_0}{2R}
\ .
\end{equation}
Expressed in terms of the same variables as the Ornstein-Uhlenbeck process, the 
correlation function of the telegraph noise model is
\begin{equation}
\label{eq: B.9}
C(t)=\frac{\sqrt{1-4\beta^2}+1}{2\sqrt{1-4\beta^2}}\exp\left[-\frac{1-\sqrt{1-4\beta^2}}{2}\gamma t\right]
      +\frac{\sqrt{1-4\beta^2}-1}{2\sqrt{1-4\beta^2}}\exp\left[-\frac{1+\sqrt{1-4\beta^2}}{2}\gamma t\right]
\ .
\end{equation}
This correlation function is significantly different from (\ref{eq: 2.2.7}); for example (\ref{eq: B.9})
is oscillatory when $\beta>1/2$.


\begin{thebibliography}{}

%
\bibitem{degennes+79}
P. G. de Gennes, 
{\sl Scaling Concepts in Polymer Physics},
Cornell University Press, (1979).
%
\bibitem{Uhl+30}
G. E. Uhlenbeck and L. S. Ornstein, 
{\sl On the theory of the Brownian motion}, 
{\it Phys. Rev.}, {\bf 36}, 823-41, (1930).
%
\bibitem{vKa81}
 N. G. van Kampen, 
{\sl Stochastic processes in Physics and Chemistry}, 2nd ed., 
North-Holland, Amsterdam, (1981).
%
\bibitem{Ber94}
M. V. Berry,
{\sl Faster than Fourier}, 
in {\sl Quantum Coherence and Reality; in celebration of the 60th Birthday of Yakir Aharonov},
(J. S. Anandan and J. L. Safko, eds.) World Scientific, Singapore, pp 55-65, (1994).
%
\bibitem{Wil87}
M. Wilkinson,
{\sl An exact effective Hamiltonian for a perturbed Landau level},
{\it J. Phys. A}. {\bf 20}, 1761-71, (1987).
%
\bibitem{Abr+72}
M. Abramowitz and I. A. Stegun (eds.), 
{\sl Handbook of Mathematical Functions},
New York: Dover, (1972).
%
\bibitem{Lan+58}
L. D. Landau and I. M. Lifshitz,
{\sl Quantum Mechanics},
Oxford: Pergamon, (1958).
%
\bibitem{Edm57}
A. R. Edmonds,
{\sl Angular Momentum in Quantum Mechanics},
Princeton, (1957).
%
\bibitem{BendOrsz+99}
C. M. Bender and S. A. Orszag, {\sl Advanced Mathematical Methods for
Scientists and Engineers: Asymptotic Methods and Perturbation Theory},
Springer-Verlag, New-York (1999).
%
\bibitem{Gut+10}
C. G\" uttler, J. Blum, A. Zsom, C. W. Ormel, and C. P. Dullemond,
{\sl The outcome of protoplanetary dust growth: pebbles, boulders,
or planetesimals? I. Mapping the zoo of laboratory collision experiments},
{\it Astron. Astrophys.}, {\bf 513}, A56, (2010).
%
\bibitem{Pru+97}
H. R. Pruppacher and J. D. Klett,
{\sl Microphysics of Clouds and Precipitation}, 2nd ed.,
Dordrecht, Kuwer, (1997).
%
\bibitem{Zim+11}
R. Zimmermann, Y. Gasteuil, M. Bourgoin, R. Volk, A. Pumir and J. F. Pinton,
{\sl Rotational intermittency and turbulence induced lift experienced by 
large particles in a turbulent flow.}
{\it Phys. Rev. Lett.}, {\bf 106}, 154501 (2011).
%
\bibitem{Zim+11b}
R. Zimmermann, Y. Gasteuil, M. Bourgoin, R. Volk, A. Pumir and J. F. Pinton,
{\sl Tracking the dynamics of thranslation and absolute oreintation of a 
sphere in a turbulent flow},
{\it Rev. Sci. Instrum. } {\bf 82}, 0333906 (2011).
%
\bibitem{Jef22}
G. B. Jeffery, 
{\sl The motion of ellipsoidal particles immersed in a viscous fluid}, 
{\it Proc. R. Soc. London, Ser. A}, {\bf 102}, 16, (1922).
%
\bibitem{Bre62}
F. P. Bretherton,
{\sl The motion of rigid particles in a shear flow at low Reynolds number},
{\it J. Fluid Mech.}, {\bf 14}, 284-304, (1962).
%
\bibitem{Gir+90}
S.S. Girimaji and S.B. Pope,
{\sl A diffusion model for velocity gradients in turbulence}, 
{\it Phys. Fluids A}, {\bf 2}, 242-56, (1990).
%
\bibitem{Bru+98}
B. K. Brunk, D. L. Koch and L. W. Lion,
{\sl Turbulent coagulation of colloidal particles},
{\it J. Fluid Mech.} (1998), {\bf 364}, 81-113, (1998).
%
\bibitem{Pum+11}
A. Pumir and M. Wilkinson,
{\sl Orientation statistics of small bodies in turbulence},
in preparation, (2011).
%
\bibitem{Sha+78}
V. E. Shapiro and V. M. Loginov,
{\sl \lq Formulae of differentiation' and their use for solving stochastic equations},
{\it Physica A}, {\bf 91}, 563-74, (1978).
%
\bibitem{Fal+07}
G. Falkovich, S. Musacchio, L. Piterbarg and M. Vucelja,
{\sl Inertial particles driven by a telegraph noise},
{\it Phys. Rev. E}, {\bf 76}, 026313, (2007).
%
\bibitem{Shi+05}
M. Shin and D. L. Koch, 
{\sl Rotational and translational dispersion of fibres
in isotropic turbulent flows},
{\it J Fluid Mech.}, {\bf 540}, 143, (2005).
%
\bibitem{Eps24}
P. S. Epstein, 
{\sl On the resistance experienced by spheres in their motion through gases},
{\it Phys. Rev.}, {\bf 22}, 710, (1924).
%
\bibitem{Par+11}
S. Parsa, J. S. Guasto, M. Kishore, N. T. Ouellette, J. P. Gollub and G. A. Voth,
{\sl Rotation and alignment of rods in two-dimensional chaotic flow},
{\it Phys. Fluids}, {\bf 23}, 043302, (2011).
%
\bibitem{Bez+10}
V. Bezuglyy, B. Mehlig and M. Wilkinson, 
{\sl Poincare indices of rheoscopic visualisations}, 
{\it Eurohys. Lett.}, {\bf 89}, 34003, (2010).
%
\bibitem{Wolfram}  Weisstein, E. W. (1999). Mathworld -- A Wolfram Web resource. URL:
http://mathworld.wolfram.com
%

\end{thebibliography}
\end{document}